\documentclass[epj]{svjour}
\usepackage{graphics}
\usepackage{epsfig,amsmath,amssymb} 
\usepackage{multirow}
\usepackage{subfig}
\usepackage{psfig}
\usepackage{color}
\begin{document}
%
\title{A frustrated quantum spin-${\boldmath s}$ model on the Union Jack lattice with spins ${\boldmath s>\frac{1}{2}}$}
\author{R.F.~Bishop
\and P.~H.~Y.~Li}
%
\offprints{}          
%
\institute{School of Physics and Astronomy, The University of Manchester, Schuster Building, Manchester, M13 9PL, United Kingdom}
%
\date{Received: date / Revised version: date}
%
%
\abstract{The zero-temperature phase diagrams of a two-dimensional
  frustrated quantum antiferromagnetic system, namely the Union Jack
  model, are studied using the coupled cluster method (CCM) for the
  two cases when the lattice spins have spin quantum number $s=1$ and
  $s=\frac{3}{2}$. The system is defined on a square lattice and the
  spins interact via isotropic Heisenberg interactions such that all
  nearest-neighbour (NN) exchange bonds are present with identical
  strength $J_{1}>0$, and only half of the next-nearest-neighbour
  (NNN) exchange bonds are present with identical strength $J_{2}
  \equiv \kappa J_{1} > 0$. The bonds are arranged such that on the $2
  \times 2$ unit cell they form the pattern of the Union Jack
  flag. Clearly, the NN bonds by themselves (viz., with $J_{2}=0$)
  produce an antiferromagnetic N\'{e}el-ordered phase, but as
  the relative strength $\kappa$ of the frustrating NNN bonds is
  increased a phase transition occurs in the classical case ($s
  \rightarrow \infty$) at $\kappa^{\rm cl}_{c}=0.5$ to a canted
  ferrimagnetic phase. In the quantum cases considered here we also
  find strong evidence for a corresponding phase transition between a
  N\'{e}el-ordered phase and a quantum canted ferrimagnetic phase at a
  critical coupling $\kappa_{c_{1}}=0.580 \pm 0.015$ for $s=1$ and
  $\kappa_{c_{1}}=0.545 \pm 0.015$ for $s=\frac{3}{2}$. In both cases
  the ground-state energy $E$ and its first derivative $dE/d\kappa$
  seem continuous, thus providing a typical scenario of a second-order
  phase transition at $\kappa=\kappa_{c_{1}}$, although the order
  parameter for the transition (viz., the average ground-state on-site
  magnetization) does not go to zero
  there on either side of the transition.  
}

\maketitle

\section{Introduction}
\label{intro}
In a recent paper~\cite{ccm_UJack_asUJ_2010} we have used the coupled
cluster method (CCM)~\cite{Bi:1991,Bi:1998,Fa:2004} to study the
magnetic order on a frustrated Heisenberg antiferromagnetic system
defined on the Union Jack lattice described in Sec.~\ref{model_sec}
below. In the earlier paper~\cite{ccm_UJack_asUJ_2010} we studied the
case of particles with spin quantum number $s=\frac{1}{2}$. In the
present paper we further the investigation of this frustrated Union
Jack model by replacing the spin-$\frac{1}{2}$ particles with particles with
higher values of $s$. In particular we study each of the cases $s=1$
and $s=\frac{3}{2}$, both of which are computationally more
challenging than the previous case with $s=\frac{1}{2}$. Just as in the
earlier paper, however, we again use the much-studied CCM. Our main
rationale for the present work is that one knows in broad terms that
the value of the spin quantum number $s$ can play an important and
often highly non-trivial role in the behaviour of strongly correlated
magnetic-lattice systems. They often exhibit rich and interesting
phase structures that are themselves a consequence of the interplay
between the quantum fluctuations and the frustration due to the
competing interactions present in the system under study. The strength
of the quantum fluctuations can itself be tuned, for example, either by
introducing anisotropy terms in the Hamiltonian (and see
Refs. \cite{Da:2004,Bi:2008_PRB,Bi:2008_JPCM} for examples) or by
varying the value of the spin quantum number $s$ of the particles (and
see Ref. \cite{Da:2005_JPhy_17} for an example). In the present paper
we study the effect on the Union Jack model of increasing $s$, in
order to throw more light on the mechanisms for magnetic ordering
inherent in the system. Alternative ways of varying the quantum
fluctuations, not studied here, would be to introduce anisotropy into
one or both of the magnetic Heisenberg bonds present, either in real
space (and see Ref. \cite{Bi:2008_JPCM} for an example) or in spin
space (and see Refs \cite{Bi:2008_PRB,Da:2005_JPhy_17} for example).

The general field of quantum magnetism at zero temperature, for spins
on the sites of regular lattices in two spatial
dimensions~\cite{2D_magnetism_1,2D_magnetism_2,2D_magnetism_3}, has
become an important and fascinating subject in recent years, and one
that is at the forefront of modern condensed matter research. Much
attention has been focussed on frustrated systems where different
types of bonds are in competition with one another, especially when
each type acting alone produces a different ground-state (gs) ordering
and a different phase. A set of touchstone problems in this respect is
the case of the two-dimensional (2D) square-lattice Heisenberg
antiferromagnet (HAF) with nearest-neighbour (NN) exchange bonds with
strength $J_{1}>0$ that by themselves produce a N\'{e}el-ordered
phase, but now with the addition of frustrating next-nearest-neighbour
(NNN) bonds of strength $J_{2} > 0$ on some or all of the diagonals of
some or all of the fundamental square-lattice plaquettes. A review
of the ``pure'' system for $J_{2} = 0$ is given in
Ref.~\cite{square_manousakis}. Although quantum fluctuations certainly
act to destroy the perfect N\'{e}el antiferromagnetic ordering of the
classical model (equivalent to the limiting case $s \rightarrow
\infty$), it is well established that for the $s=\frac{1}{2}$ case the
order parameter, namely the sublattice or staggered magnetization, has
a value equal to about 61\% of the classical limiting value. For such
unfrustrated 2D models the most accurate results are generally
provided by quantum Monte Carlo (QMC) simulations (and see, e.g.,
Ref.~\cite{Sandvik:1997} for the spin-half HAF on the 2D
square lattice).

Once NNN bonds are added to the above pure spin-$\frac{1}{2}$ HAF the situation
becomes much more complicated. The prototypical model in this respect
is the so-called $J_{1}$--$J_{2}$ model in which all possible NNN bonds
are included, and which finds good experimental realization in such
layered materials as Li$_{2}$VOSiO$_{4}$, BaCdVO(PO$_{4}$)$_{2}$, and
others. Various approximate methods have been used to simulate the
properties of this system, including the coupled cluster method (CCM)
\cite{Bi:2008_PRB,Bi:2008_JPCM,j1j2_square_ccm1,j1j2_square_ccm4,j1j2_square_ccm5},
series expansion (SE) techniques
\cite{j1j2_square_series1,j1j2_square_series2,j1j2_square_series3,j1j2_square_series4,j1j2_square_series5},
exact diagonalization (ED) methods
\cite{j1j2_square_ed1,j1j2_square_ed2,j1j2_square_ed3}, and
hierarchical mean-field (MF) calculations~\cite{j1j2_square_mf}. For
frustrated spin-lattice models in two dimensions both the QMC and ED
techniques face formidable difficulties. These arise in the former
case due to the ``minus-sign problem'' present for frustrated systems
when the nodal structure of the gs wave function is unknown, and in
the latter case due to the practical restriction to relatively small
lattices imposed by computational limits. The latter problem is
exacerbated for incommensurate phases, and is compounded due to the
large (and essentially uncontrolled) variation of the results with
respect to the different possible shapes of clusters of a given size.

Several other models of this same general class of spin-$\frac{1}{2}$
models on the 2D square lattice with both NN and NNN interactions
present, but in which some of the NNN $J_{2}$ bonds are removed, have
prompted considerable recent interest. One such is the
Shastry-Sunderland model~\cite{shastry1,shastry2,shastry3}, in which
three-quarters of the $J_{2}$ bonds of the $J_{1}$--$J_{2}$ model are
removed in a particular arrangement so that each lattice site is
connected by four NN $J_{1}$ bonds and one NNN $J_{2}$ bond. The
Shastry-Sunderland model finds a good experimental realization in the
magnetic material SrCu(BO$_{3}$)$_{2}$. A second such model, in which
half of the $J_2$ bonds of the $J_{1}$--$J_{2}$ model are removed in
an arrangement that leaves each lattice site connected by four NN
$J_{1}$ bonds and two NNN $J_{2}$ bonds, is the anisotropic triangular
lattice (or $J_{1}$-$J_{2}'$) model~\cite{square_triangle}, which is
also believed to well describe the layered magnetic material
Cs$_{2}$CuCl$_{4}$. A third such model, in which again half of the
$J_{2}$ bonds of the $J_{1}$--$J_{2}$ model are removed, but now in an
arrangement that leaves half of the sites eight-connected (namely, by
four NN $J_{1}$ bonds and four NNN $J_{2}$ bonds) and the other half
four-connected by $J_{1}$ bonds alone, is the Union Jack
model~\cite{ccm_UJack_asUJ_2010,Co:2006_PRB,Co:2006_JPhys,Zh:2007},
whose study we continue here by extending the situation to where the
spins have spin quantum number $s>\frac{1}{2}$. Although all of the
models mentioned above show antiferromagnetic N\'eel ordering for
small values of $J_2$, their phase diagrams for larger values of $J_2$
display a wide variety of behaviour, including, two-dimensional
quantum ``spirals'', valence-bond crystals/solids, and spin
liquids~\cite{Balent:2010}. Thus, in the absence of any definitive
overarching theoretical argument, the best way to understand this
class of NN/NNN models on the square lattice is to treat each one on a
case-by-case basis.

While the general trend is that as the spin quantum number $s$ is
increased the effects of quantum fluctuations reduce, one also knows
that there can be significant deviations from it.  A particularly
well-known example is the since-confirmed prediction of Haldane that
integer-spin systems on the linear chain would have a nonzero
excitation energy gap, whereas half-odd-integer spin systems would be
gapless~\cite{Ha:1983}.  Indeed, such deviations from general trends
provide one of the main reasons why quantum spin-lattice problems
still maintain such an important role in the general investigation of
quantum phase transitions.

For the past few decades, a great deal of attention has also been
devoted to magnetic materials with spin-1 ions (see, e.g.,
Refs.~\cite{Bi:2008_JPCM_anisotropy_XXZ_2D} and references cited
therein). In this context we note the recent discovery of
superconductivity with a transition temperature at $T_c \approx 26\,$K
in the layered iron-based compound LaOFeAs, when doped by partial
substitution of the oxygen atoms by fluorine atoms~\cite{KWHH:2008},
La[O$_{1-x}$F$_x$]FeAs, with $x \approx$ 0.05--0.11.  This has been
followed by the rapid discovery of superconductivity at even higher
values of $T_c$ ($\gtrsim 50\,$K) in a broad class of similar doped
quaternary oxypnictide compounds.  Enormous interest has thereby been
engendered in this class of materials. Of particular relevance to the
present work are the very recent first-principles
calculations~\cite{MLX:2008} showing that the undoped parent precursor
material LaOFeAs is well described by the spin-1 $J_{1}$--$J_{2}$
model on the square lattice with $J_1 > 0$, $J_2 > 0$, and
$J_{2}/J_{1} \approx 2$. Broadly similar conclusions have also been
reached by other authors~\cite{SA:2008}. It is clear, therefore, that
the theoretical study of 2D quantum magnets with $s=1$ (or $s > 1$)
are worthy of pursuit in their own right, as well as to highlight
differences with their $s=\frac{1}{2}$ counterpart.

In this article we continue the study, begun in
Ref.~\cite{ccm_UJack_asUJ_2010} for the $s=\frac{1}{2}$ case, of
magnetic ordering in the Union Jack model, by now extending the
discussion to the two cases where the spins have either $s=1$ or
$s=\frac{3}{2}$. The model itself is discussed in more detail in
Sec.~\ref{model_sec} below. For the $s=\frac{1}{2}$ case the model has
been studied previously using SWT~\cite{Co:2006_PRB,Co:2006_JPhys} and
SE techniques~\cite{Zh:2007}, as well as by
the CCM~\cite{ccm_UJack_asUJ_2010} used here. However, to our knowledge,
no results have yet been reported for the cases with
$s>\frac{1}{2}$. 

As in the case of the spin-half anisotropic triangular lattice (or
$J_{1}$--$J_{2}'$) model, it was shown
\cite{Co:2006_PRB,Co:2006_JPhys,Zh:2007} that NN N\'eel order for the
$s=\frac{1}{2}$ Union Jack model persists until a critical value of
the frustrating NNN ($J_2$) bonds. However, in contrast to the case of
the anisotropic triangular lattice model, there exists a ferrimagnetic
ground state in which spins on the eight-connected sites cant at a
nonzero angle with respect to their directions in the corresponding
N\'{e}el state. This model thus exhibits an overall magnetic moment in
this regime, which is quite unusual for spin-$\frac{1}{2}$ 2D materials with
only Heisenberg bonds and which therefore preserve (spin) rotational
symmetries in the Hamiltonian. This model also presents us with a
difficult computational task in order to simulate its properties. Here
we wish to continue to study this model for the $s=1$ and $s=3/2$
cases using the CCM, which has consistently been shown to yield
insight into a wide range of problems in quantum magnetism.

\section{The model}
\label{model_sec}
\begin{figure}[t]
\begin{center}
\mbox{
   \subfloat[Canted\label{canted_model}]{\scalebox{0.45}{\epsfig{file=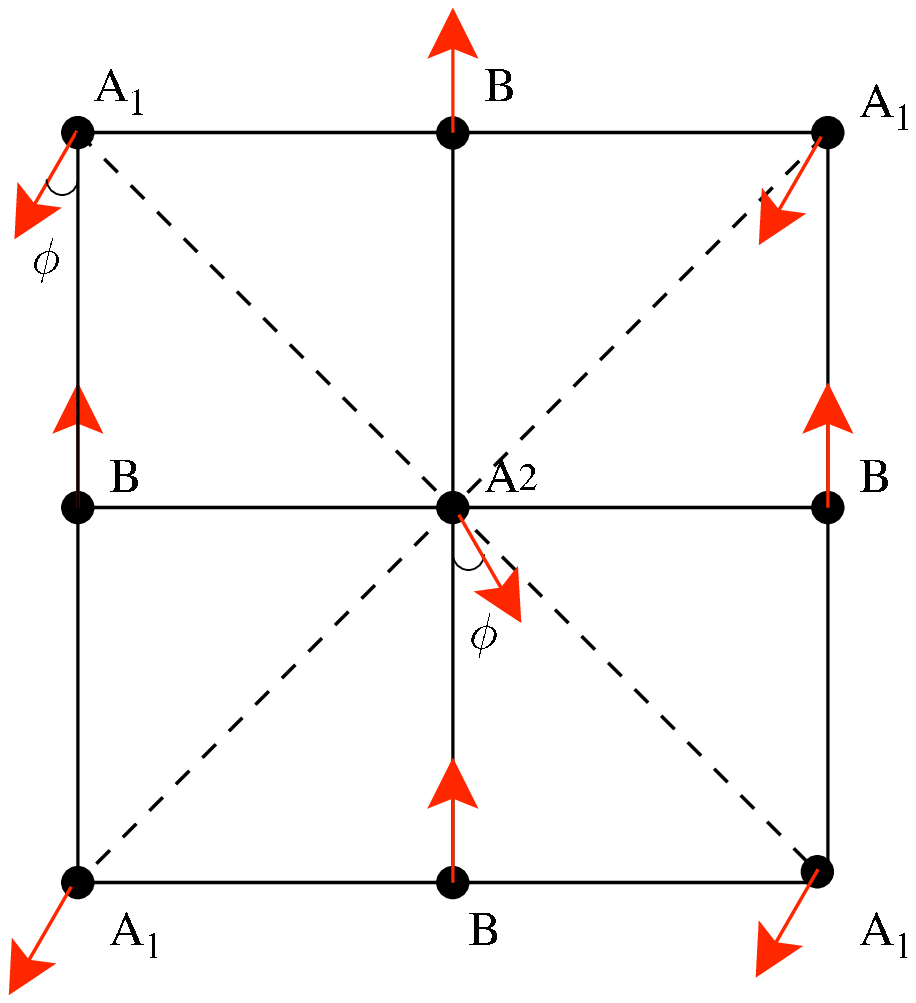}}}
   \subfloat[Semi-striped\label{SemiStripe_model}]{\scalebox{0.45}{\epsfig{file=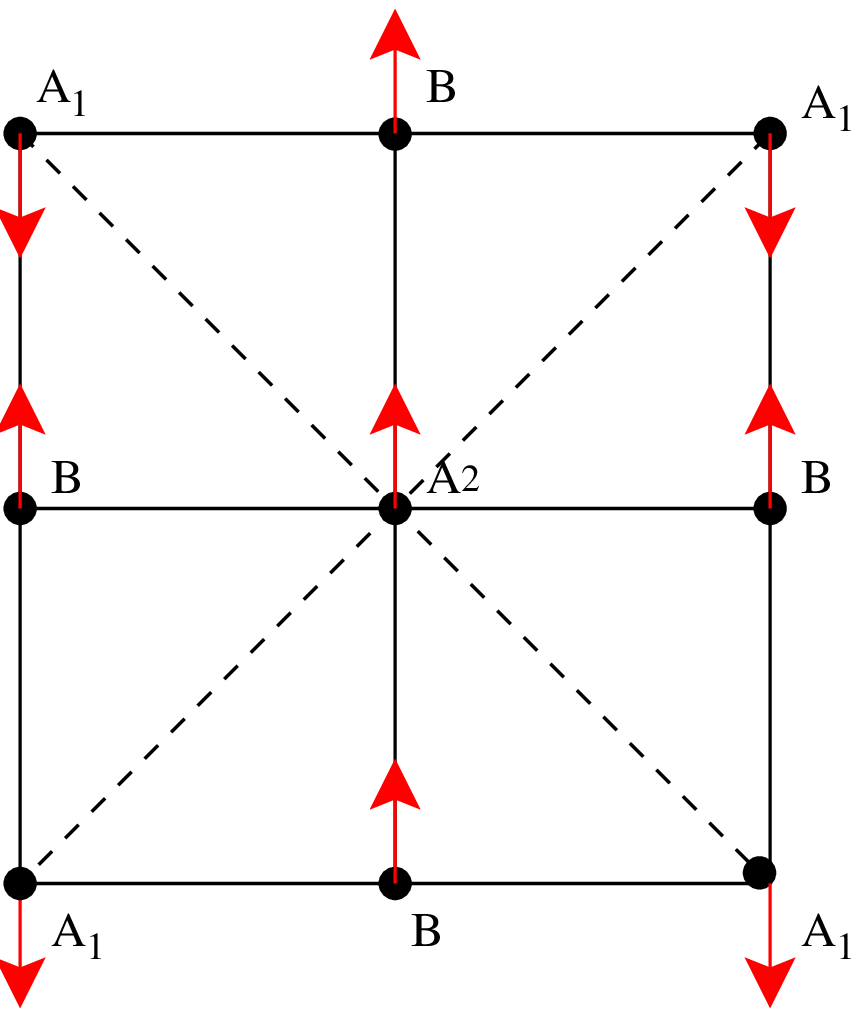}}}
}
\caption{Union Jack model; --- $J_{1}$; - - -
  $J_{2}$. (a) Canted state; (b) semi-striped state. The unit cell is a
  square of side length 2.}
\label{model}
\end{center}
\end{figure}

The Hamiltonian of the Union Jack model considered here is given by
\begin{equation}
H = J_{1}\sum_{\langle i,j \rangle}{\bf s}_{i}\cdot{\bf s}_{j} + J_{2}\sum_{[i,k]}{\bf s}_{i}\cdot {\bf s}_{k}\,,  \label{H}
\end{equation}
where the operators ${\bf s}_{i} \equiv (s^{x}_{i}, s^{y}_{i},
s^{z}_{i})$ are the quantum spin operators on lattice site $i$ with
${\bf s}^{2}_{i} = s(s+1)$ where, for the cases considered here, $s=1,\frac{3}{2}$. On the underlying 2D square
lattice the sum over $\langle i,j \rangle$ runs over all distinct NN
bonds with strength $J_{1}$, while the sum over [$i,k$] runs over only
half of the distinct NNN diagonal bonds having strength $J_{2}$ and
with only one diagonal bond on each square plaquette as arranged in
the pattern shown explicitly in Fig.~\ref{model}.
The unit cell is thus the $2 \times 2$ square shown in
Fig.~\ref{model}. (We note that, by contrast, the $J_{1}$--$J_{2}$
model discussed above includes all of the diagonal NNN bonds on the
square lattice, and its unit cell is thus the $1 \times 1$ square.) We
again consider the case where both sorts of bonds are
antiferromagnetic, $J_{1} > 0$ and $J_{2} \equiv \kappa J_{1} > 0$,
and thus act to compete against (or to frustrate) each
other. Henceforth we set $J_{1} \equiv 1$. We consider the model
equivalently defined by the Union Jack geometry in which there are two
sorts of sites, namely the A sites with eight NN sites and the B sites
with four NN sites, as shown in Fig.~\ref{model}(a).

Considered as a classical model, (and thus corresponding to the
quantum case in the limit where the spin quantum number $s \rightarrow
\infty$), the Union Jack model has only two ground-state (gs) phases
as the parameter $\kappa$ is varied over the range (0, $\infty$). A
simple variational analysis for the classical model reveals that for
$0 < \kappa$ $< \frac{1}{2}$ the gs phase is N\'{e}el-ordered, exactly as for
the full $J_{1}$--$J_{2}$ model. The N\'{e}el ordering induced by the
$J_{1}$ bonds acting alone is hence preserved as the strength of the
competing $J_{2}$ bonds is increased, until the critical value
$\kappa^{{\rm cl}}_{c}=0.5$ is reached. For $\kappa > \kappa^{{\rm
    cl}}_{c}$ a new phase of lower energy emerges, just as in the full
$J_{1}$--$J_{2}$ model. For the full $J_{1}$--$J_{2}$ model that new
phase is a classical striped state in which alternate rows (or
columns) of spins are arranged antiparallel to one another, whereas
the new classical gs phase for the Union Jack model is the canted
ferrimagnetic state shown in Fig.~\ref{model}(a). In the canted
state the spins on each of the alternating A$_{1}$ and A$_{2}$ sites
of the A-sublattice are canted respectively at angles ($\pi \mp \phi$)
with respect to those on the B sublattice, all of the latter of which
point in the same direction. On the A sublattice each site A$_{1}$ has
four NN sites A$_{2}$, and vice versa. The angle between the NN spins
on the A sublattice is thus 2$\phi$.

The energy of the above canted state for the classical model is thus 
\begin{equation}
E=Ns^{2}(\kappa\,{\rm cos}\,2\phi - 2\,{\rm cos}\,\phi)\,,
\end{equation}
where $J_{1} \equiv 1$ and $N \rightarrow \infty$ is the number of
sites. The energy is thus extremized when
\begin{equation}
{\rm sin}\,\phi(1 - 2 \kappa\, {\rm cos}\, \phi) = 0\,.
\end{equation}
When $\kappa < \kappa^{{\rm cl}}_{c} \equiv 0.5$, the lowest energy
corresponds to sin$\,\phi=0$ and hence to the N\'{e}el state. By
contrast, when $\kappa > \kappa^{{\rm cl}}_{c} \equiv 0.5$ the lowest
energy solution is the canted state with
\begin{equation}
\phi_{{\rm cl}} = {\rm cos}^{-1}\,\bigg(\frac{1}{2\kappa}\bigg) \,.  \label{pitch_angle}
\end{equation}
Thus the classical gs energy is given by
\begin{equation}
E^{{\rm cl}} = \left\{ 
\begin{array}{l l}
  Ns^{2}(\kappa -2)\,; & \quad \kappa < \kappa^{{\rm cl}}_{c} \equiv 0.5 \,.\\
  Ns^{2}\big(-\frac{1}{2\kappa} - \kappa\big)\,; & \quad \kappa > \kappa^{{\rm cl}}_{c} \equiv 0.5\,.\\ \end{array} \right. 
\end{equation}
The classical phase transition at $\kappa = \kappa^{{\rm cl}}_{c}
\equiv 0.5$ is of continuous (second-order) type with the gs energy
and its first derivative with respect to $\kappa$ both continuous functions of $\kappa$,
although there are finite discontinuities in the second- and
higher-order derivatives at $\kappa = \kappa^{{\rm cl}}_{c}$.

The total magnetization per site in the canted phase for the classical model is
$m^{{\rm cl}}=\frac{1}{2}s[1-(2\kappa)^{-1}]$, and the model thus exhibits
ferrimagnetism in this phase. Whereas ferrimagnetism more commonly
occurs when the individual ionic spins have different magnitudes on
different sublattices, it arises here in a case where when the spins
all have the same magnitude and all the interactions are
antiferromagnetic in nature, but the frustration between them acts to
produce an overall magnetization. The total magnetization $m$ vanishes
linearly as $\kappa \rightarrow \kappa^{{\rm cl}}_{c}$ from the canted
phase and then remains zero in the N\'{e}el phase for $\kappa <
\kappa$$^{{\rm cl}}_{c}$. The spontaneous breaking of the spin rotation
symmetry is also reflected by the vanishing of the energy gap on both
sides of the transition. Clearly on both sides of the transition the
translation symmetry of the lattice is also broken.

As previously for the $s=\frac{1}{2}$ case~\cite{ccm_UJack_asUJ_2010}, our aim here is to give a fully microscopic
analysis of the Union Jack model for the quantum case where the spins
all have spin quantum number either $s=1$ or $s=\frac{3}{2}$. Our goal is to
map out the zero-temperature ($T=0$) phase diagrams for both cases,
including the positions and orders of any quantum phase transitions
that emerge. In particular we investigate the quantum analogues of the
classical N\'{e}el and canted phases and calculate the effect of
quantum fluctuations on the position and nature of the transition
between them. We also aim to investigate, for particular regions of
the control parameter $\kappa$, whether the quantum fluctuations may
favour other phases, which have no classical counterparts. One such
possible candidate is discussed below.

For the classical ($s \rightarrow \infty$) model the $\kappa
\rightarrow \infty$ limit corresponds to a canting angle $\phi \rightarrow
\frac{1}{2} \pi$, such that the spins on the A sublattice become
N\'{e}el-ordered, as is expected. The spins on the
antiferromagnetically-ordered A sublattice are thus orientated at
90$^{\circ}$ to those on the ferromagnetically-ordered B sublattice in
this limit. In reality, of course, there is complete degeneracy at the
classical level in this limit between all states for which the
relative ordering directions for spins on the A and B sublattices are
arbitrary. Clearly the exact $\kappa \rightarrow \infty$ limit of the
Union Jack model with spins having any value of the spin quantum
number $s$ should also comprise decoupled antiferromagnetic and
ferromagnetic sublattices. However, one might now expect that this
degeneracy in the relative spin orientations between the two
sublattices is lifted by quantum fluctuations by the well-known
phenomenon of {\it order by disorder}~\cite{Vi:1977}. Just such a
phase is known to exist in the full spin-$\frac{1}{2}$
$J_{1}$--$J_{2}$ model for values of $J_{2}/J_{1} \gtrsim 0.6$, where
it is the so-called collinear striped phase in which, on the square
lattice, spins along (say) the rows in Fig.\ \ref{model} order
ferromagnetically while spins along the columns and diagonals order
antiferromagnetically. We have also shown how such a striped state is
stabilized by quantum fluctuations for values of $J_{2}'/J_{1} \gtrsim
1.8$ for the spin-$\frac{1}{2}$ anisotropic triangular lattice model
(or $J_{1}$--$J_{2}'$ model)~\cite{square_triangle}.

The existence of the striped state as a stable phase for large values
of the frustration parameter for both the spin-$\frac{1}{2}$
$J_{1}$--$J_{2}$ and $J_{1}$--$J_{2}'$ models above is a reflection of
the well-known fact that quantum fluctuations favour collinear
ordering. In both cases the order-by-disorder mechanism favours the
collinear state from the otherwise infinitely degenerate set of
available states at the classical level. For the present Union Jack
model the corresponding collinear state that might perhaps be favoured
by the order by disorder mechanism is the so-called semi-striped state
shown in Fig.~\ref{model}(b) where the A sublattice is now
N\'{e}el-ordered in the same direction as the B sublattice is
ferromagnetically ordered. Alternate rows (or columns) are thus
ferromagnetically and antiferromagnetically ordered in the same
direction. We investigate the possibility below that if such a
semi-stripe-ordered phase may be stabilized by quantum fluctuations at
larger values of $\kappa$ for either of the cases $s=1$ or
$s=\frac{3}{2}$, in order to compare with the earlier $s=\frac{1}{2}$
case~\cite{ccm_UJack_asUJ_2010}.

We note that for the $s=\frac{1}{2}$ case our own CCM
calculations~\cite{ccm_UJack_asUJ_2010} provided strong evidence that
the canted ferrimagnetic phase becomes unstable at large values of the
frustration parameter $\kappa$. In view of that observation we also
used the CCM for the $s=\frac{1}{2}$ case with the collinear
semi-stripe-ordered ferrimagnetic state as a model state. We found
tentative evidence, based on the relative energies of the two states,
for a second zero-temperature phase transition between the canted and
semi-stripe-ordered ferrimagnetic states at a larger critical value of
$\kappa_{c_{2}} \approx 125 \pm 5$, as well as firm evidence for a
first phase transition between the N\'{e}el antiferromagnetic phase
and the canted ferrimagnetic phase at a critical coupling
$\kappa_{c_{1}} = 0.66 \pm 0.02$. Our prediction for
$\kappa_{c_{2}}$, however, was based on an extrapolation of the CCM
results for the canted state into regimes where the solutions have
already become unstable and the CCM equations based on the canted
state as model state have no solutions at any level of (LSUB$n$)
approximation beyond the lowest (with $n=2$). The prediction for
$\kappa_{c_{2}}$ was thus less reliable than that for
$\kappa_{c_{1}}$, although our results showed clear evidence that, if
the second transition at $\kappa = \kappa_{c_{2}}$ does exist, it
should be of first-order type. By contrast, the transition at $\kappa
= \kappa_{c_{1}}$ for the $s=\frac{1}{2}$ case was found to be an
interesting one. As in the classical ($s \rightarrow \infty$) case,
the energy and its first derivative were seen to be continuous
(within the errors inherent in our approximations), thus providing a
typical scenario of a second-order phase transition, although a weakly
first-order one could not be excluded. Nevertheless, the average
on-site magnetization was seen to approach a nonzero value $M_{c_{1}}
= 0.195 \pm 0.005$ on both sides of the transition, which is more
typical of a first-order transition. The slope, $dM/d\kappa$, of the
order parameter curve as a function of the coupling parameter
$\kappa$, also appeared to be continuous, or very nearly so, at the
critical point $\kappa_{c_{1}}$. Thus, all of the evidence shows that
for the $s=\frac{1}{2}$ case the transition between the N\'{e}el and
canted phases is a subtle one. A particular interest here is to
compare and contrast the corresponding transition(s) between the
$s=\frac{1}{2}$ and the $s>\frac{1}{2}$ models.

We first briefly describe the main elements of the CCM below in
Sec.~\ref{CCM}, where we also discuss the approximation schemes used
in practice for the $s=\frac{1}{2}$ case and the $s>\frac{1}{2}$
cases. Then in Sec.~\ref{Results} we present our CCM results based on
using the N\'{e}el, canted and semi-striped states discussed above as
model states (or starting states). We conclude in
Sec.~\ref{discussion} with a discussion of the results.

\section{The coupled cluster method}
\label{CCM}
It is widely recognized nowadays that the CCM (see, e.g.,
Refs.~\cite{Bi:1991,Bi:1998,Fa:2004} and references cited therein)
employed here is one of the most powerful and most versatile modern
techniques in quantum many-body theory.  It has been successfully
applied to various quantum magnets (see, e.g.,
Refs.~\cite{Fa:2004,Bi:2008_PRB,Bi:2008_JPCM,j1j2_square_ccm1,j1j2_square_ccm4,j1j2_square_ccm5,square_triangle,Fa:2001,Kr:2000,Schm:2006},
and is particularly appropriate for studying frustrated systems, for
which the main alternative methods are often only of limited
usefulness.  For example, QMC techniques are particularly plagued by
the sign problem for such systems, and the ED method is often
restricted in practice, particularly for $s>1/2$, to such small
lattices that it is often insensitive to the details of any subtle
phase order present.

We now briefly describe the CCM means to solve the ground-state (gs)
Schr\"{o}dinger ket and bra equations, $H|\Psi\rangle = E|\Psi\rangle$
and $\langle\tilde{\Psi}|H=E\langle\tilde{\Psi}|$ respectively (and
see Refs.~\cite{Bi:1991,Bi:1998,Fa:2004,Fa:2001,Kr:2000,Schm:2006} for
further details). The first step in implementing the CCM is always to
choose a normalized starting state or model state $|\Phi\rangle$ on top of which
to incorporate later in a systematic fashion the multi-spin
correlations contained in the exact ground states $|\Psi\rangle$ and
$\langle\tilde{\Psi}|$. More specifically, the CCM employs the
exponential parametrizations, $|\Psi\rangle=e^{S}|\Phi\rangle$ and
$\langle\tilde{\Psi}|=\langle\Phi|\tilde{\cal S}$e$^{-S}$. These
states are chosen with a normalization such that $\langle \tilde{\Psi}|\Psi
\rangle = 1$ [i.e., with $\langle {\tilde\Psi}|=(\langle \Psi|\Psi
\rangle)^{-1}\langle\Psi|$], and with $|\Psi\rangle$ itself satisfying
the intermediate normalization condition $\langle \Phi|\Psi \rangle =
1 = \langle \Phi|\Phi \rangle$. The correlation operator $S$ is
expressed as $S = \sum_{I\neq0}{\cal S}_{I}C^{+}_{I}$ and its
counterpart is $\tilde{S} = 1 + \sum_{I\neq0}\tilde{\cal
  S}_{I}C^{-}_{I}$. The operators $C^{+}_{I} \equiv
(C^{-}_{I})^{\dagger}$, with $C^{+}_{0} \equiv 1$, have the property
that $\langle\Phi|C^{+}_{I} = 0 = C^{-}_{I}|\Phi\rangle;\, \forall I
\neq 0$. They form a complete set of multi-spin creation operators with
respect to the model state $|\Phi\rangle$. The index $I$ is a
set-index that stands for the set of lattice sites whose spin
projections on the quantization axis are changed with respect to their
values in the model state $|\Phi\rangle$. The ket- and bra-state
correlation coefficients $({\cal S}_{I}, \tilde{{\cal S}_{I}})$ are
calculated by requiring the gs energy expectation value $\bar{H}
\equiv \langle\tilde{\Psi}|H|\Psi\rangle$ to be a minimum with respect
to each of them.  This immediately yields the coupled set of equations
$\langle \Phi|C^{-}_{I}\mbox{e}^{-S}H\mbox{e}^{S}|\Phi\rangle = 0$ and
$\langle\Phi|\tilde{S}(\mbox{e}^{-S}H\mbox{e}^{S} -
E)C^{+}_{I}|\Phi\rangle = 0\,;\, \forall I \neq 0$, which we solve in
practice for the correlation coefficients $({\cal S}_{I}, \tilde{{\cal
    S}_{I}})$ within specific truncation schemes described below, by
making use of parallel computing routines~\cite{ccm}.

It is important to note that while the above CCM parametrizations of
$|\Psi\rangle$ and $\langle{\tilde \Psi}|$ are not manifestly Hermitian
conjugate, they do preserve the important Hellmann-Feynman theorem at
{\it all} levels of approximation (viz., when the complete set of
many-particle configurations $\{I\}$ is
truncated~\cite{Bi:1998}. Furthermore, the amplitudes (${\cal
  S}_{I},{\tilde{\cal S}}_{I}$) form canonically conjugate pairs in a
time-dependent version of the CCM, in contrast with the pairs (${\cal
  S}_{I},{\cal S}^{\ast}_{I}$) that come from a manifestly Hermitian
conjugate representation for
$\langle{\tilde\Psi}|=(\langle\Phi|e^{S^{\dagger}}e^{S}|\Phi\rangle)^{-1}\langle\Phi|e^{S^{\dagger}}$,
which are {\it not} canonically conjugate to each
other~\cite{Bi:1998}.

In order to treat each lattice site on an equal footing we perform a
mathematical rotation of the local spin axes on each lattice site, such
that every spin of the model state aligns along its negative $z$-axis.
Henceforth our description of the spins is given wholly in terms of
these locally defined spin coordinate frames.  In particular, the
multi-spin creation operators may be written as \(C^{+}_{I}\equiv
s^{+}_{i_{1}} s^{+}_{i_{2}} \cdots s^{+}_{i_{n}}\), in terms of the
locally defined spin-raising operators $s^{+}_{i} \equiv s^{x}_{i} +
s^{y}_{i}$ on lattice sites $i$.  Having solved for the multi-spin
cluster correlation coefficients $({\cal S}_{I}, \tilde{{\cal
    S}_{I}})$ as described above, we may then calculate the gs energy
$E$ from the relation
$E=\langle\Phi|\mbox{e}^{-S}H\mbox{e}^{S}|\Phi\rangle$, and the
average gs on-site magnetization $M$ from the relation $M \equiv
-\frac{1}{N} \langle\tilde{\Psi}|\sum_{i=1}^{N}s^{z}_{i}|\Psi\rangle$
which holds in the rotated local spin coordinates. The quantity $M$ is
thus the magnetic order parameter, and it is just the usual sublattice
(or staggered) magnetization for the case of the N\'{e}el state as CCM
model state, for example.

Although the CCM formalism is clearly exact if a complete set of multi-spin
configurations $\{I\}$ with respect to the model state $|\Phi\rangle$ 
is included in the calculation of the correlation operators
$S$ and $\tilde{S}$, in practice it is necessary to use systematic approximation 
schemes to truncate them to some finite subset. In our earlier paper on the $s=\frac{1}{2}$ 
version of the present model~\cite{ccm_UJack_asUJ_2010}, 
we employed, as in our previous
work~\cite{Fa:2004,Kr:2000,Schm:2006,Ze:1998,Fa:2002},
the localized LSUB$n$ scheme 
in which all possible multi-spin-flip correlations over different locales
on the lattice defined by $n$ or fewer contiguous lattice sites are retained.

However, we note that the number of fundamental LUB$n$ configurations
for $s=1$ and $s=\frac{3}{2}$ becomes appreciably higher than for
$s=\frac{1}{2}$, since each spin on each site $i$ can now be flipped
up to 2$s$ times by the spin-raising operator $s^{+}_{i}$. Thus, for
the cases of $s=1$ and $s=\frac{3}{2}$ it is more practical, but
equally systematic, to use the alternative SUB$n$-$m$ scheme, in which
all correlations involving up to $n$ spin flips spanning a range of no
more than $m$ adjacent lattice sites are
retained~\cite{Fa:2004,Fa:2001}. We then set $m=n$, and hence employ
the so-called SUB$n$--$n$ scheme. More generally, the LSUB$m$ scheme
is thus equivalent to the SUB$n$-$m$ scheme for $n=2sm$ for particles
of spin $s$. For $s=\frac{1}{2}$, LSUB$n\equiv$ SUB$n$-$n$; whereas
for $s=1$, LSUB$n\equiv$ SUB2$n$-$n$, and for $s=\frac{3}{2}$, ${\rm
  LSUB}n = {\rm SUB}3n$-$n$. The numbers of such fundamental
configurations (viz., those that are distinct under the symmetries of
the Hamiltonian and of the model state $|\Phi\rangle$) that are
retained for the N\'{e}el and semi-striped states of the current $s=1$
and $s=\frac{3}{2}$ models at various SUB$n$--$n$ levels are shown in
Table~\ref{FundConf_SUBnn}. We note that the distinct configurations
given in Table~\ref{FundConf_SUBnn} are defined with respect to the
Union Jack geometry described in Sec.~\ref{model_sec}, in which the B
sublattice sites of Fig.~\ref{model}(a) are defined to have four NN
sites, and the A sublattice sites are defined to have the eight NN
sites joined to them by either $J_{1}$ bonds or $J_{2}$ bonds. If we
chose instead to work in the square-lattice geometry, by contrast,
each site would have four NN sites.
\begin{table}[t]
  \caption{Number of fundamental SUB$n$-$n$ configurations ($N_{f}$) for the semi-striped and 
    canted states of the spin-$1$ and spin-$\frac{3}{2}$ Union Jack models, based on the Union Jack geometry defined in the text.}
\label{FundConf_SUBnn}     
\vskip0.1cm
\begin{tabular}{cccccc} \hline\hline \\ [-7pt]
\multirow{3}{*}{Method} &  \multicolumn{2}{c}{$s=1$} & & \multicolumn{2}{c}{$s=\frac{3}{2}$}   \\  \cline{2-3} \cline{5-6}  \\   [-7pt]   
 &  \multicolumn{2}{c}{$N_{f}$}  & & \multicolumn{2}{c}{$N_{f}$}  \\ \cline{2-3} \cline{5-6}  \\ [-7pt]
& semi-striped  & canted   & & semi-striped  & canted \\ \hline \\ [-7pt]
SUB2-2 &  3    & 9 &  & 3 &   9   \\ 
SUB4-4 & 115   & 556 &  &  115 &  618  \\ 
SUB6-6 & 7826  &  52650 &  & 9862 & 68365  \\ \hline\hline
\end{tabular} 
\end{table}

Although we never need to perform any finite-size scaling, since all
CCM approximations are automatically performed from the outset in the
$N \rightarrow \infty$ limit, we do need as a last step to extrapolate
to the $n \rightarrow \infty$ limit in the truncation index $n$.  We
use the same well-tested scaling laws as for the $s=\frac{1}{2}$ model for the
gs energy per spin $E/N$ and the average gs on-site magnetization $M$,
\begin{equation}
E/N=a_{0}+a_{1}n^{-2}+a_{2}n^{-4},  \label{Extrapo_E}
\end{equation} 
\begin{equation}
M=b_{0}+b_{1}n^{-1}+b_{2}n^{-2}. \label{Extrapo_M}
\end{equation} 

\section{Results}
\label{Results}
We present results of CCM calculations for both the spin-1 and
spin-$\frac{3}{2}$ Union Jack models with the Hamiltonian of Eq.\
(\ref{H}), for given parameters ($J_{1}=1$, $J_{2}$), based
respectively on the N\'{e}el, canted and semi-striped states as CCM
model states. Our computational power is such that we can perform
SUB$n$-$n$ calculations for each model state with $n \leq 6$.

\subsection{N\'{e}el state versus the canted state}
Results are first presented that are obtained using the N\'{e}el and
canted model states. While classically we have a second-order phase
transition from N\'{e}el order (for $\kappa < \kappa$$^{{\rm
    cl}}_{c}$) to canted order (for $\kappa > \kappa^{{\rm cl}}_{c}$),
where $\kappa \equiv J_{2}/J_{1}$, at a value $\kappa^{{\rm cl}}_{c} =
0.5$, using the CCM we find strong indications of a shift of this
critical point to a higher value $\kappa_{c_{1}} \approx 0.58$ in the
spin-1 case and $\kappa_{c_{1}} \approx 0.545$ in the
spin-$\frac{3}{2}$ quantum case as we explain in detail below. These
may be compared with the even higher value $\kappa_{c_{1}} \approx
0.66$ found previously~\cite{ccm_UJack_asUJ_2010} for the
spin-$\frac{1}{2}$ case.

\begin{figure*}[!htb]
\mbox{
  \subfloat[$s=1$]{\scalebox{0.3}{\includegraphics[angle=270]{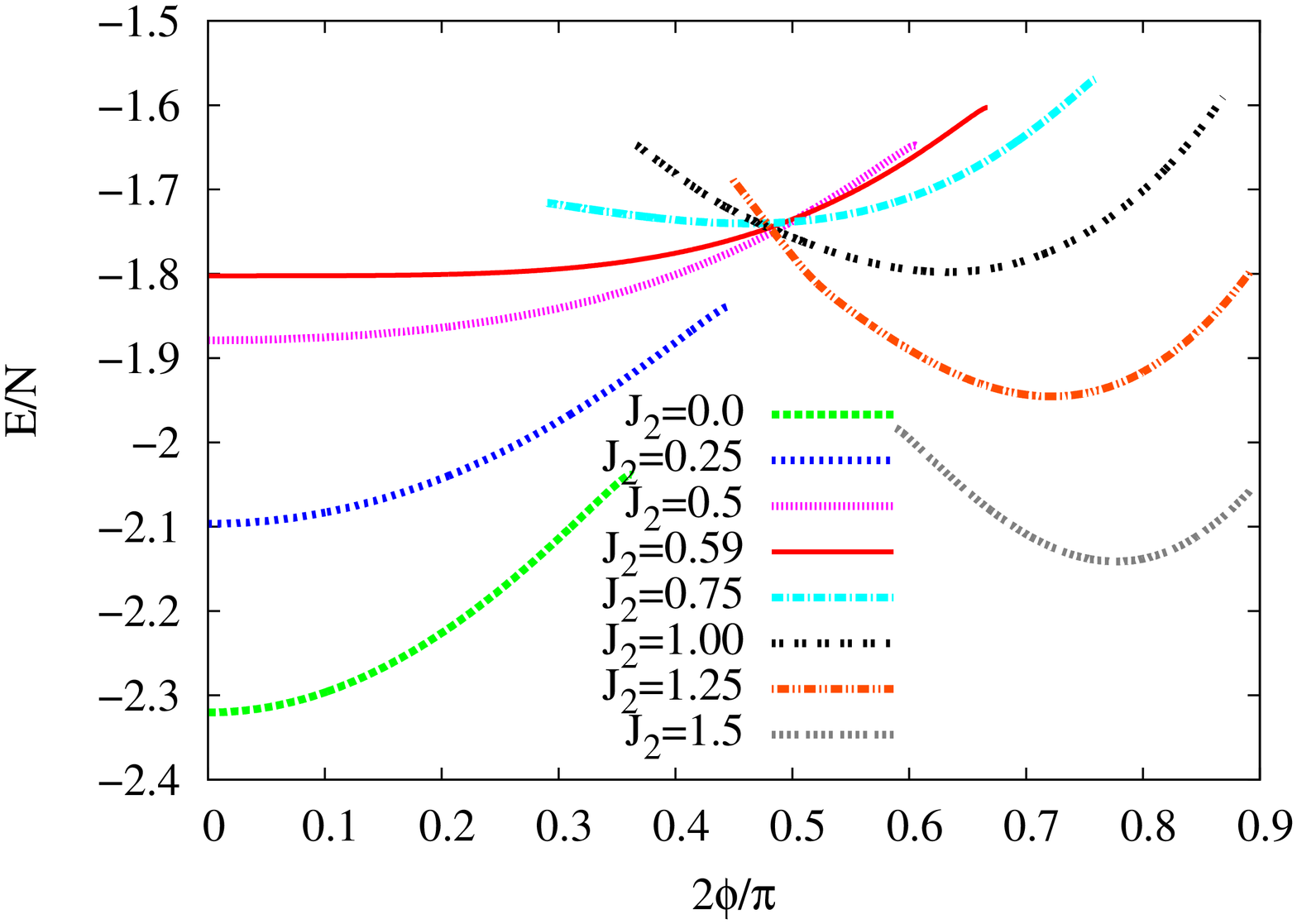}}}
  \subfloat[$s=3/2$]{\scalebox{0.3}{\includegraphics[angle=270]{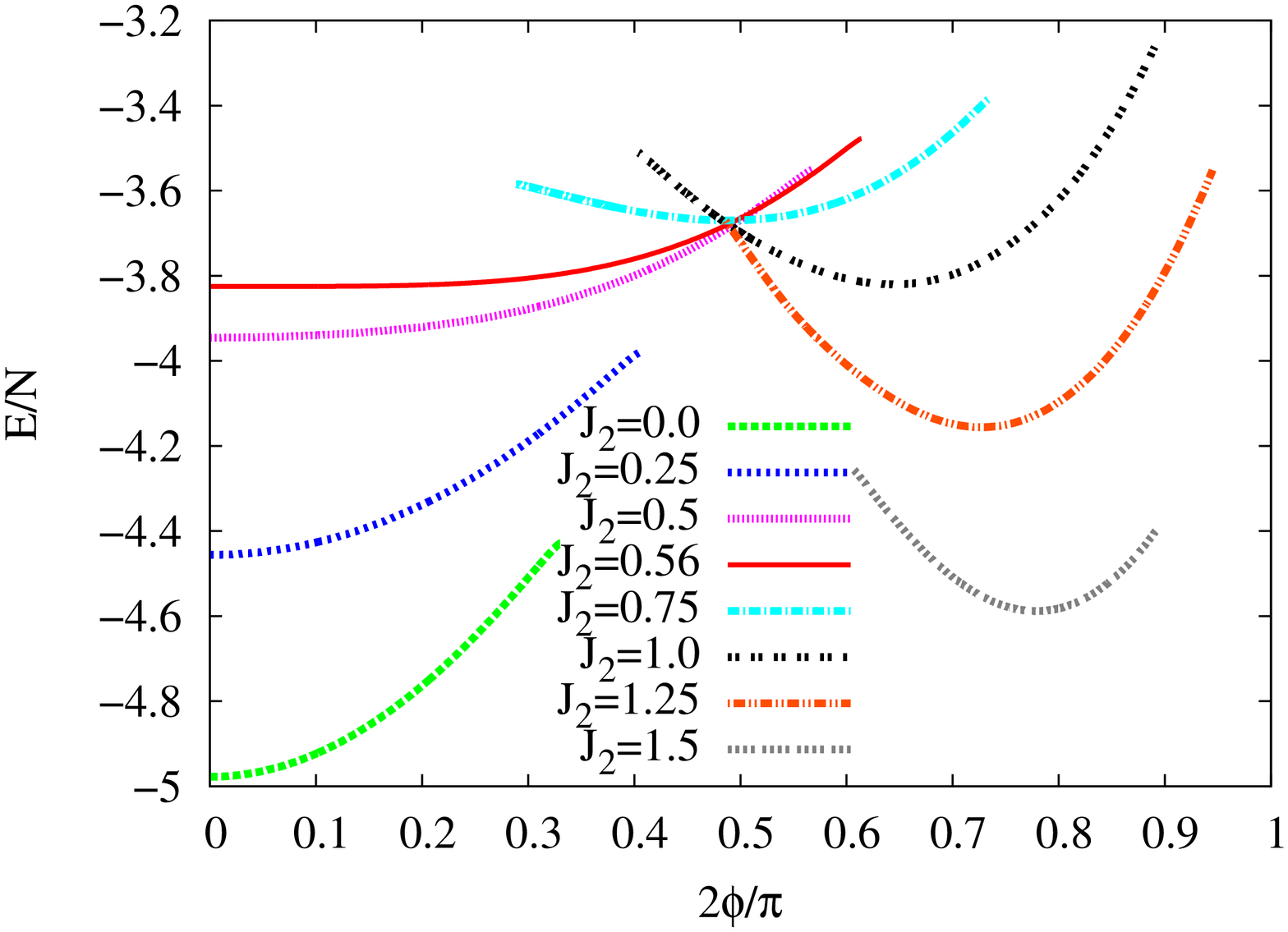}}}
}
\mbox{
  \subfloat[$s=1$]{\scalebox{0.3}{\includegraphics[angle=270]{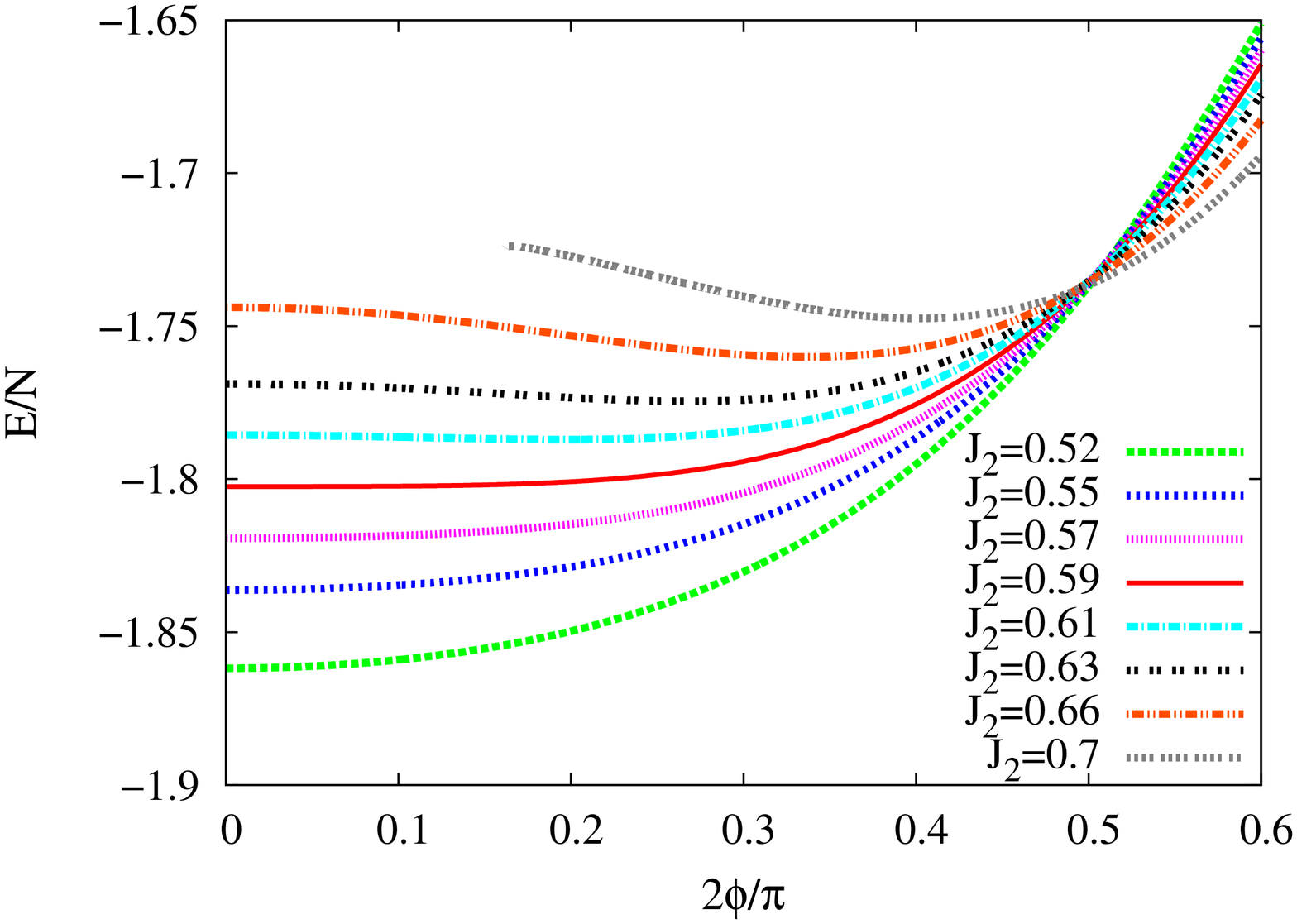}}}
  \subfloat[$s=3/2$]{\scalebox{0.3}{\includegraphics[angle=270]{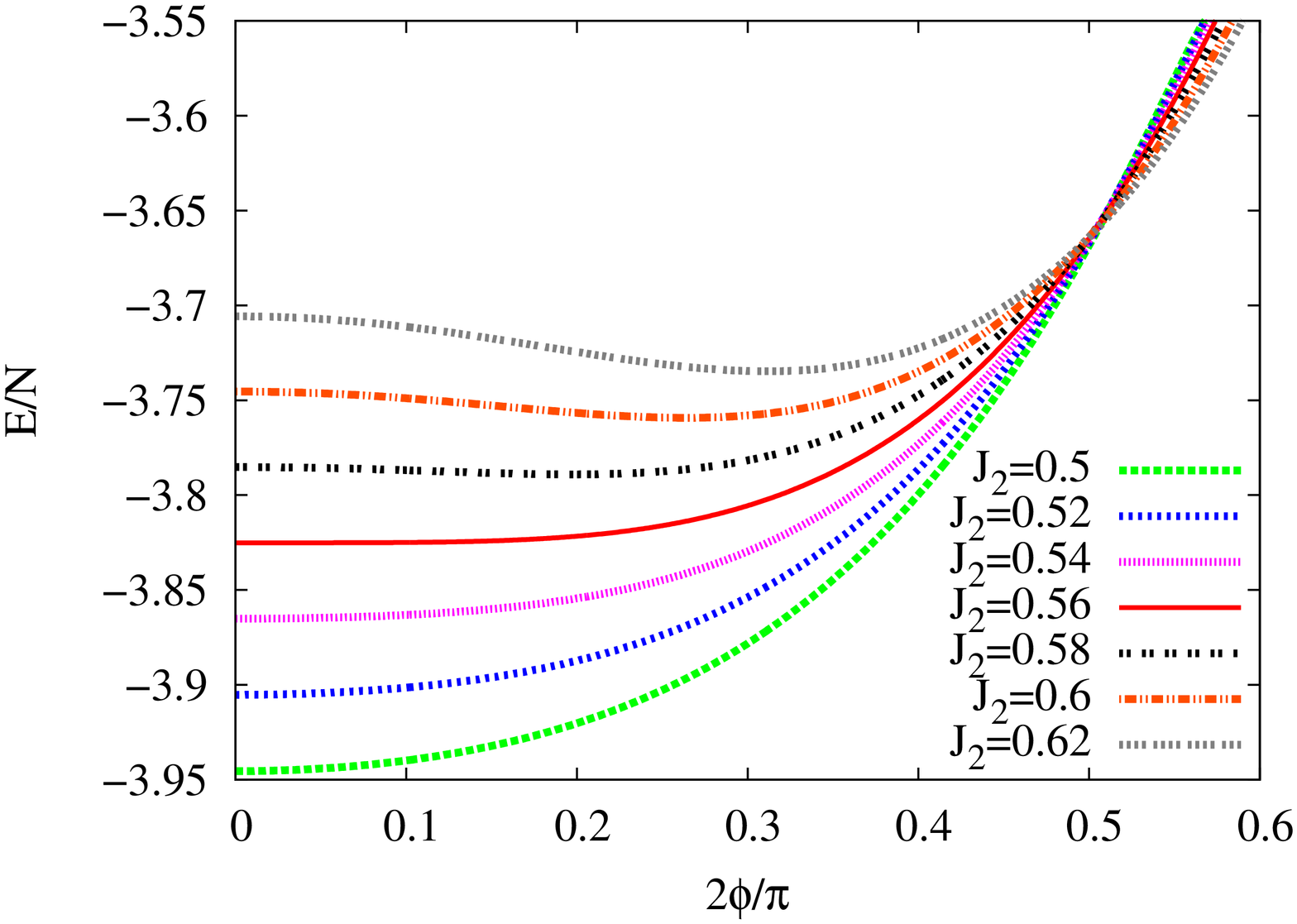}}}
}
\caption{Ground-state energy per spin of the spin-1 and
  spin-$\frac{3}{2}$ Union Jack Hamiltonian of Eq.\ (\ref{H}) with
  $J_{1} \equiv 1$, using the SUB4-4 approximation of the CCM with the
  canted model state, versus the canting angle $\phi$, for various
  values of $J_{2}$. For the case of $s=1$, for $J_{2} \lesssim 0.592$
  in this approximation the minimum is at $\phi=0$ (N\'{e}el order),
  whereas for $J_{2} \gtrsim 0.592$ the minimum occurs at
  $\phi=\phi_{{\rm SUB}4-4} \neq 0$, indicating a phase transition at
  $J_{2} \approx 0.592$ in this SUB4-4 approximation. For the case of
  $s=\frac{3}{2}$, for $J_{2} \lesssim 0.560$ in this approximation
  the minimum is at $\phi=0$ (N\'{e}el order), whereas for $J_{2}
  \gtrsim 0.560$ the minimum occurs at $\phi=\phi_{{\rm SUB}4-4} \neq
  0$, indicating a phase transition at $J_{2} \approx 0.560$ in this
  SUB4-4 approximation.}
\label{EvsAngle}
\end{figure*}

\begin{figure*}[!htb]
\mbox{
    \subfloat[$s=1$]{\scalebox{0.3}{\includegraphics[angle=270]{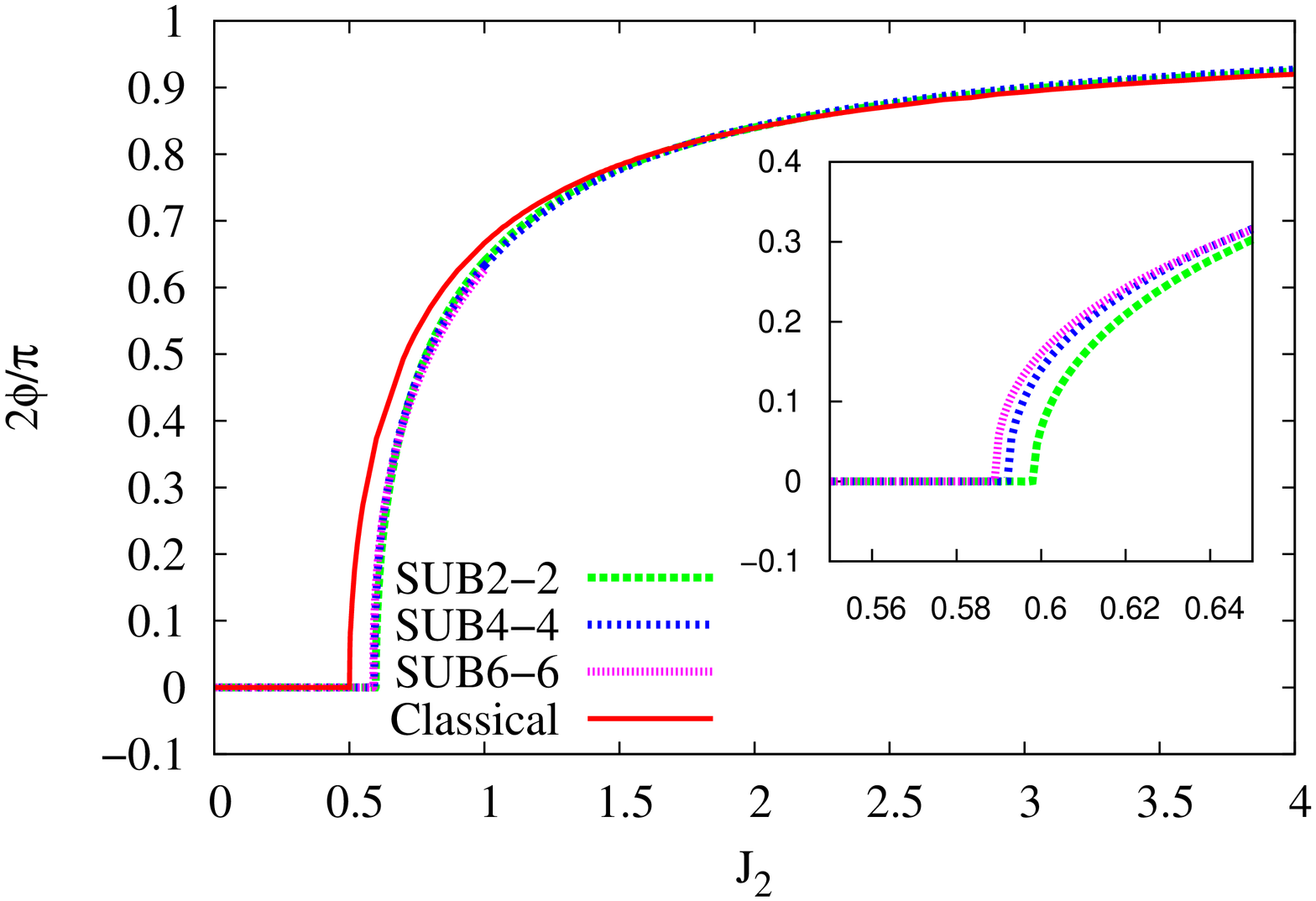}}}
    \subfloat[$s=3/2$]{\scalebox{0.3}{\includegraphics[angle=270]{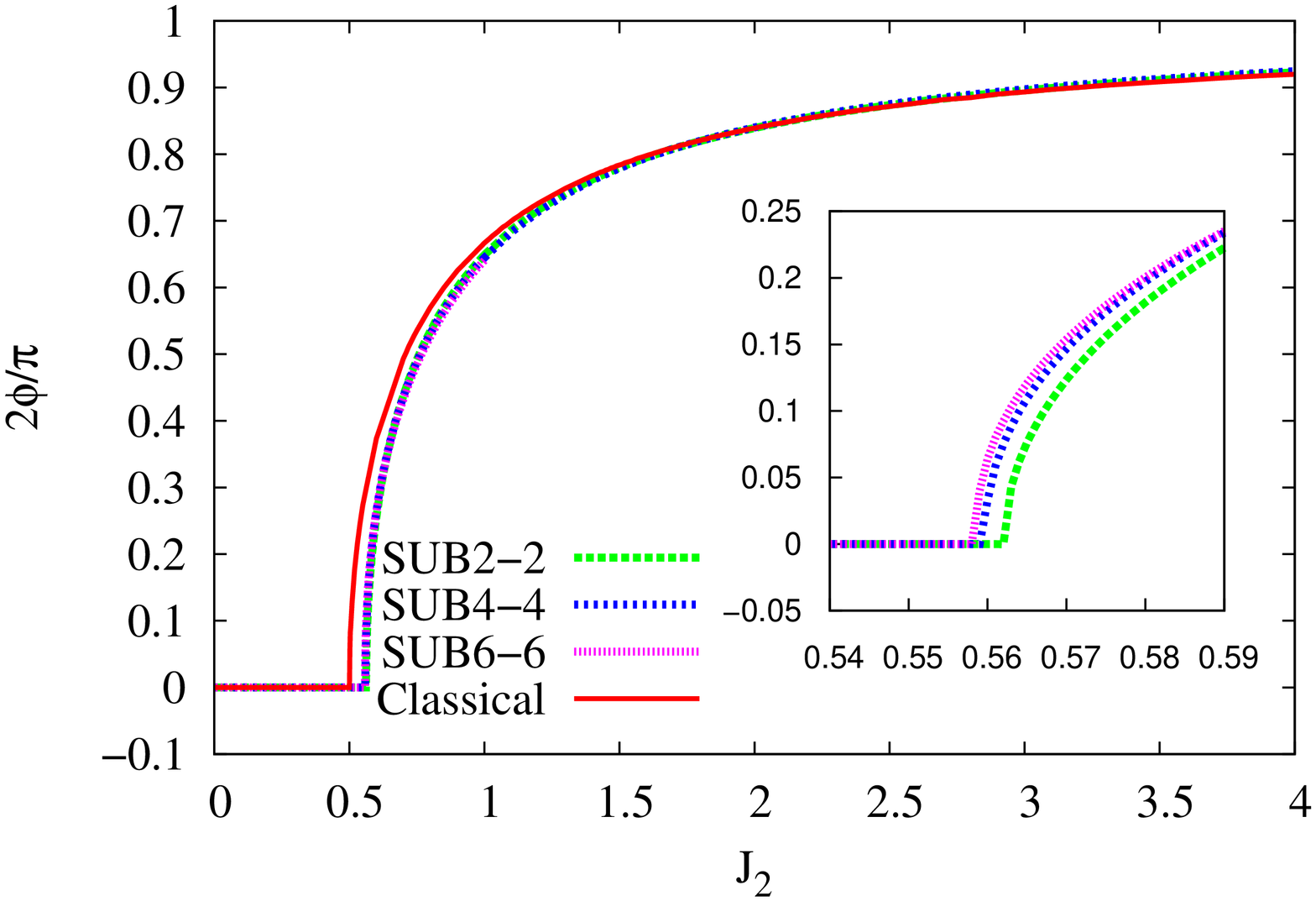}}}
}
\caption{The angle $\phi_{{\rm SUB}n-n}$ that minimizes the energy
  $E_{{\rm SUB}n-n}(\phi)$ of the (a) spin-1 and (b)
  spin-$\frac{3}{2}$ Union Jack Hamiltonian of Eq.\ (\ref{H}) with
  $J_{1} \equiv 1$, in the SUB$n$-$n$ approximations with
  $n=\{2,4,6\}$, using the canted model state, versus $J_{2}$. The
  corresponding classical result $\phi_{{\rm cl}}$ from
  Eq.~(\ref{pitch_angle}) is shown for comparison. We find in the
  SUB$n$-$n$ quantum case with $n \geq 2$ a second-order phase transition
  (e.g., for SUB6-6 at $J_{2} \approx 0.589$ for $s=1$ and $J_{2}
  \approx 0.559$ for $s=\frac{3}{2}$). In the classical case there is
  a second-order phase transition at $J_{2}=0.5$.}
\label{angleVSj2}
\end{figure*}

Thus, just as for the spin-$\frac{1}{2}$ case, curves such as those
in Fig.~\ref{EvsAngle}
show that the N\'{e}el model state ($\phi=0$) gives the minimum gs
energy for all values of $\kappa < \kappa_{c_{1}}$ where
$\kappa_{c_{1}} = \kappa^{{\rm SUB}n-n}_{c_{1}}$ is also dependent on
the level of SUB$n$-$n$ approximation, as we see clearly in Fig.\
\ref{angleVSj2}.
Conversely, for $\kappa > \kappa_{c_{1}}$ the minimum in the energy
is found to occur at a value $\phi \neq 0$. If we consider the canting
angle $\phi$ itself as an order parameter (i.e., $\phi=0$ for N\'{e}el
order and $\phi \neq 0$ for canted order) a typical scenario for a
first-order phase transition would be the appearance of a two-minimum
structure for the gs energy as a function of $\phi$. If we therefore
admit such a scenario, in the typical case one would expect various
special points in the transition region, namely the phase transition
point $\kappa_{c_{1}}$ itself where the two minima have equal depth,
plus one or two instability points $\kappa_{i_{1}}$ and
$\kappa_{i_{2}}$ where one or other of the minima (at $\phi = 0$ and
$\phi \neq 0$ respectively) disappears. By contrast, a second-order
phase transition might manifest itself via a one-minimum structure for
the gs energy as a function of $\phi$, in which the single minimum
moves smoothly and continuously from the value $\phi = 0$ for all
values of $\kappa < \kappa_{c_{1}}$ to nonzero value $\phi \neq 0$ for
$\kappa > \kappa_{c_{1}}$. 

Results for the gs energy per spin calculated in the SUB4-4
approximation based on the canted state as the CCM model state are
shown in Fig.~\ref{EvsAngle} as a function of the canting
angle. Curves such as those shown in Fig.~\ref{EvsAngle} for the
SUB4-4 case show that what happens for this model at this level of
approximation is that for $\kappa \lesssim 0.592$ for $s=1$ and for
$\kappa \lesssim 0.560$ for $s=\frac{3}{2}$, the only minimum in the
gs energy is at $\phi = 0$ (N\'{e}el order). As these values are
approached from below the SUB4-4 energy curves become extremely flat
near $\phi = 0$, indicating the disappearance at $\phi = 0$ of the
second derivative $d^{2}E/d\phi^{2}$ (and possibly also of one or more
of the higher derivatives $d^{n}E/d\phi^{n}$ with $n \geq$ 3), as well
as of the first derivative $dE/d\phi$. Then, for all values $\kappa
\gtrsim 0.592$ for $s=1$ and $\kappa \gtrsim 0.560$ for
$s=\frac{3}{2}$, the SUB4-4 curves develop a minimum at a value $\phi
\neq 0$ which is also the global minimum. The state for $\phi \neq 0$
is thus the quantum analogue of the classical canted phase. The fact
that the antiferromagnetic N\'{e}el order survives into the
classically unstable regime is another example of the well-known
phenomenon that quantum fluctuations tend to promote collinear order
in magnetic spin-lattice systems, as has been observed in many other
such cases (see e.g., Ref.~\cite{Kr:2000,Ko:1996}). Thus, this
collinear N\'{e}el-ordered state survives into a region where
classically it becomes unstable with respect to the non-collinear
canted state. As expected, as $s$ is increased the value of
$\kappa_{c_{1}}$ decreases towards the classical value 0.5

A detailed inspection of the curves shown in Fig.~\ref{angleVSj2} for
various SUB$n$-$n$ approximation shows that the crossover from one
minimum ($\phi = 0$, N\'{e}el) solution to the other ($\phi \neq 0$,
canted) appears to be continuous for both the $s=1$ and $s=\frac{3}{2}$ cases
indicating a second-order transition according to the above
scenario. Thus, based on the evidence presented so far of the gs energies of the
N\'{e}el and canted phases, it would appear that the transition at
$\kappa = \kappa_{c_{1}}$ between these two phases is
second-order. 

Table~\ref{table_CritPt}
\begin{table}[!b]
  \caption{The critical value $\kappa^{{\rm SUB}n-n}_{c_{1}}$ at which the transition between the N\'{e}el phase ($\phi=0$) and the canted phase ($\phi \neq 0$) occurs in the SUB$n$--$n$ approximation using the CCM with (N\'{e}el or) canted state as model state.}
\label{table_CritPt}    
\vskip0.1cm
\begin{center}
\begin{tabular}{cccc} \hline\hline  \\ [-7pt]
\multirow{2}{*}{Method} & $s=1$ & & $s=\frac{3}{2}$ \\  \cline{2-2} \cline{4-4}   \\ [-7pt]
 & $\kappa^{{\rm SUB}n-n}_{c_{1}}$ & & $\kappa^{{\rm SUB}n-n}_{c_{1}}$ \\ \hline   \\ [-7pt]
SUB2-2 & 0.598 & &  0.563  \\ 
SUB4-4 & 0.592 & &  0.560  \\ 
SUB6-6 & 0.589 & &  0.559  \\ 
SUB$\infty$ & $0.585 \pm 0.002$ & & $0.557 \pm 0.001$ \\ \hline\hline
\end{tabular}     
\end{center}
\end{table}
shows the critical values $\kappa^{{\rm SUB}n-n}_{c_{1}}$ at which the
transition between the N\'{e}el and canted phases occurs in the
various SUB$n$-$n$ approximations shown in Fig.~\ref{angleVSj2}. In
the past we have found that a simple linear extrapolation,
$\kappa^{{\rm SUB}n-n}_{c_{1}} = a_{0}+a_{1}n^{-1}$, yields a good fit
to such critical points. This seems to be the case here too, just as
for the spin-$\frac{1}{2}$ case~\cite{ccm_UJack_asUJ_2010}. The
corresponding ``SUB$\infty$'' estimates from the SUB$n$-$n$ data in
Table~\ref{table_CritPt} are $\kappa_{c_{1}}=0.585 \pm 0.002$ for
$s=1$, and $\kappa_{c_{1}}=0.557 \pm 0.001$ for $s=\frac{3}{2}$, where
the quoted errors are simply a combination of the standard deviations
from the fits and of the computational uncertainties associated with
the $\kappa^{{\rm SUB}n-n}_{c_{1}}$ points themselves. We also present
other independent estimates of $\kappa_{c_{1}}$ below.

Figure~\ref{EvsAngle} also displays the feature that for certain
values of $J_{2}$ with $J_{1} \equiv 1$ (or, equivalently, $\kappa$)
CCM solutions at a given SUB$n$-$n$ level of approximation (viz.,
SUB4-4 in Fig.\ \ref{EvsAngle}) exist only for certain ranges of the
canting angle $\phi$. For example, for the pure square-lattice HAF
($\kappa=0$) the CCM SUB4-4 solution based on a canted model state
only exists for $0 \leq \phi \lesssim 0.182 \pi$ when $s=1$, and for
$0 \leq \phi \lesssim 0.166 \pi$ when $s=\frac{3}{2}$. In this case, where the
N\'{e}el solution is the stable ground state, if we attempt to move
too far away from N\'{e}el collinearity the CCM equations themselves
become ``unstable'' and simply do not have a real solution. Similarly,
we see from Fig.~\ref{EvsAngle} that for $\kappa=1.5$, for example,
the CCM SUB4-4 solution exists only for $0.294 \pi \lesssim \phi \leq
0.446 \pi$ when $s=1$ and for $0.302 \pi \lesssim \phi \leq 0.446 \pi$ when
$s=\frac{3}{2}$. In this case the stable ground state is a canted phase, and
now if we attempt either to move too close to N\'{e}el collinearity or
to increase the canting angle too close to its asymptotic value of
$\pi/2$, the real solution terminates.

Terminations of CCM solutions like those discussed above are very
common and have been very well documented~\cite{Fa:2004}. Such termination
points always arise due to the solution of the CCM equations becoming
complex there. Beyond such points there exist two branches of entirely
unphysical complex conjugate solutions~\cite{Fa:2004}. In the region
where the solution reflecting the true physical solution is real there
actually also exists another (unstable) real solution. However, only
the shown branch of these two solutions reflects the true (stable)
physical ground state, while the other branch does not. The physical
branch is usually easily identified in practice as the one which
becomes exact in some known (e.g., perturbative) limit. It then meets
the corresponding unphysical branch at some termination point (with
infinite slope on Fig.\ \ref{EvsAngle}), beyond which no real
solutions exist. The LSUB$n$ or SUB$n$-$n$ termination points are
themselves also reflections of the quantum phase transitions in the
real system, and may be used to estimate the position of the phase
boundary~\cite{Fa:2004}, although we do not do so for this first
critical point since we have more accurate criteria, one of which has
already been discussed above. Another will be discussed below.

Before doing so, however, we wish to give some further indication of
the accuracy of our results. Thus in Table~\ref{table_EandM_results}
\begin{table}[!htb]
  \caption{Ground-state energy per spin $E/N$ and magnetic order parameter $M$ (i.e., the average on-site magnetization) for the spin-1 and spin-$\frac{3}{2}$ square-lattice HAF. We show CCM results obtained for the Union Jack model with $J_{1}=1$ and $J_{2}=0$ using the N\'{e}el model state in various CCM SUB$n$-$n$ approximations defined on the Union Jack geometry described in Sec.~\ref{model_sec}. We compare our extrapolated ($n \rightarrow \infty$) results using Eqs.~(\ref{Extrapo_E}) and (\ref{Extrapo_M}) and our SUB$n$-$n$ data sets with other calculations.}
\begin{tabular}{cccccc} \hline\hline  \\ [-7pt]
\multirow{2}{*}{Method} & \multicolumn{2}{c}{$s=1$} & & \multicolumn{2}{c}{$s=\frac{3}{2}$} \\  \cline{2-3} \cline{5-6}  \\ [-7pt]
  &  $E/N$ & $M$ & & $E/N$ & $M$   \\  \hline  \\ [-7pt]
SUB2-2 & -2.29504 &  0.9100 & & -4.94393 & 1.4043  \\     
SUB4-4 & -2.32033 &  0.8687 & &-4.97758 & 1.3618   \\      
SUB6-6 & -2.32537 &  0.8488 & & -4.98352 & 1.3423   \\          
SUB$\infty$   & -2.3295  & 0.800 & & -4.9882 & 1.295  \\  
SWT $^{a}$ & -2.3282  &  0.8043 & & & \\ 
SE $^{b}$  &  -2.3279(2)  & 0.8039(4) & & &  \\ \hline\hline
\end{tabular} 
\vspace{0.3cm}
\\ 
\protect $^{a}$ SWT (Spin Wave Theory) for square lattice~\cite{Ha:1992_SWT_Square_s1}  \\ 
\protect $^{b}$ SE (Series Expansion) for square lattice~\cite{Zh:1991_SE_Square_s1} 
\label{table_EandM_results}
\end{table}
we show data for the cases of the spin-1 and spin-$\frac{3}{2}$ HAF on
the square lattice (corresponding to the case $\kappa=0$ of the
present Union Jack model). We present our CCM results in various
SUB$n$-$n$ approximations (with $2 \leq n \leq 6$) based on the Union
Jack geometry using the N\'{e}el model state. Results are given for
the gs energy per spin $E/N$, and the magnetic order parameter $M$,
which in this case is simply the staggered magnetization. We also
display our extrapolated ($n \rightarrow \infty$) results using the
schemes of Eqs.~(\ref{Extrapo_E}) and (\ref{Extrapo_M}) with the data
set $n=\{2,4,6\}$. The results are robust, and for
comparison purposes we also show for the spin-1 case, the
corresponding results using a spin wave theory (SWT) technique
\cite{Ha:1992_SWT_Square_s1} and from a linked-cluster series
expansion (SE) method~\cite{Zh:1991_SE_Square_s1}. Our own extrapolated
results are in good agreement with these results and own previous CCM
results~\cite{Fa:2001_Square_s1}.

\begin{figure*}[!htb]
\mbox{
 \subfloat[$s=1$\label{E_spin1}]{\scalebox{0.3}{\includegraphics[angle=270]{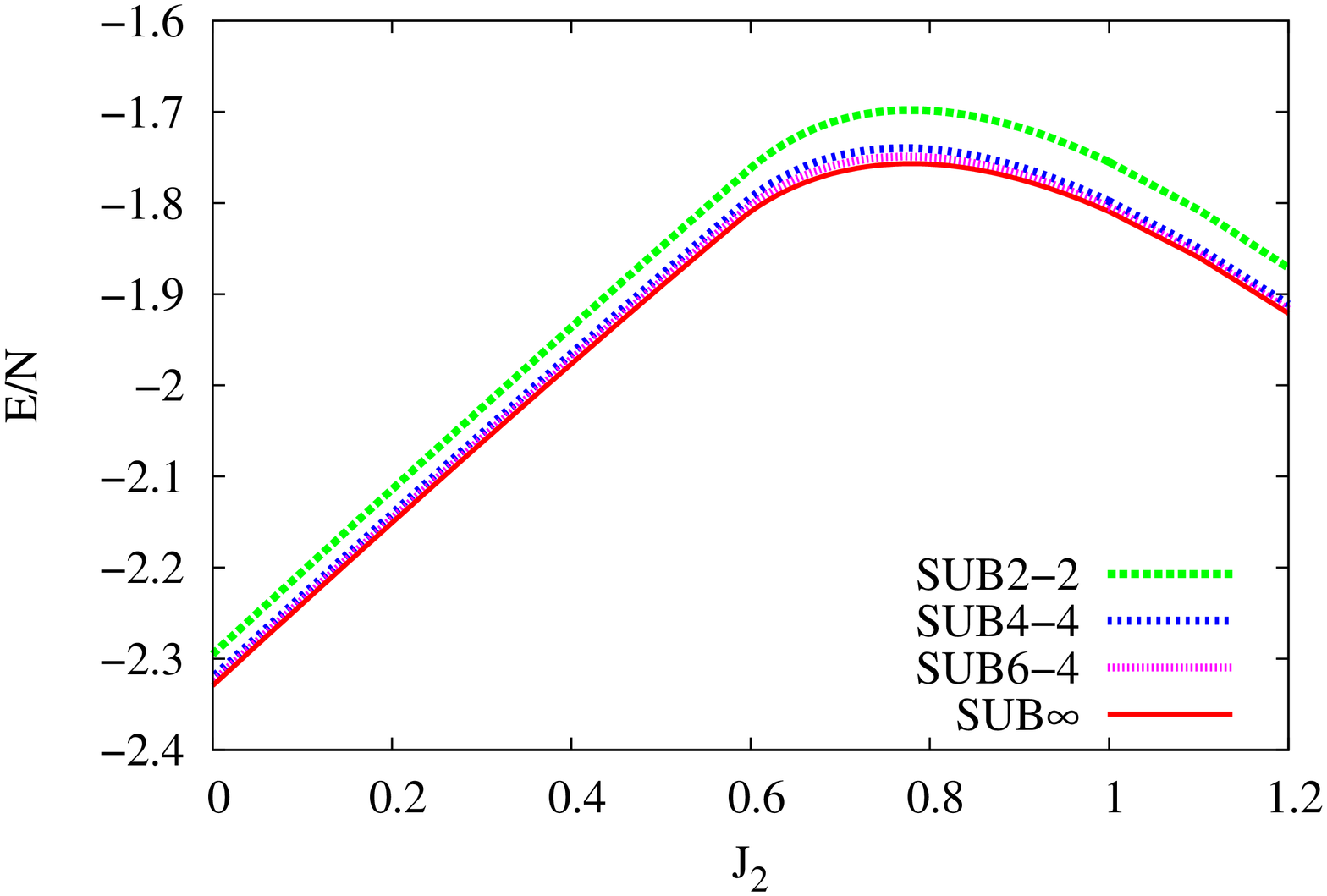}}}
 \subfloat[$s=3/2$\label{E_spin1.5}]{\scalebox{0.3}{\includegraphics[angle=270]{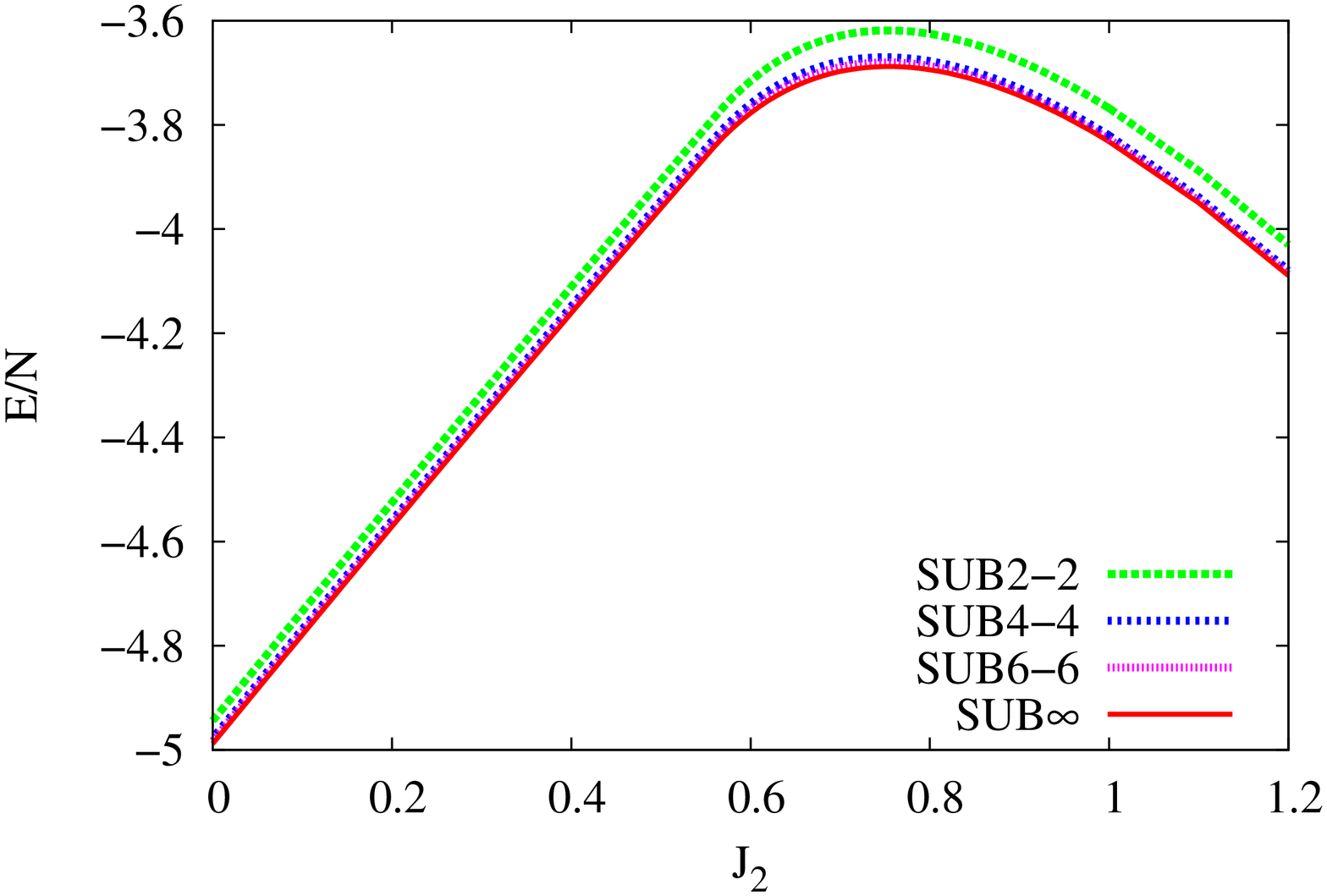}}}
}
\caption{Ground-state energy per spin versus $J_{2}$ for the N\'{e}el
  and canted phases of the (a) spin-1 and (b) spin-$\frac{3}{2}$ Union
  Jack Hamiltonian of Eq.\ (\ref{H}) with $J_{1} \equiv 1$. The CCM
  results using the canted model state are shown for various SUB$n-n$
  approximations with $n=\{2,4,6\}$ with the canting angle
  $\phi=\phi_{{\rm LSUB}n}$ that minimizes $E_{{\rm SUB}n-n}(\phi)$.
  We also show the $n \rightarrow \infty$ extrapolated result from
  using Eq.\ (\ref{Extrapo_E}).}
\label{E}
\end{figure*}

The CCM results for the gs energy per spin are shown in Fig.~\ref{E}
for various SUB$n$--$n$ approximations based on the canted (and
N\'{e}el) model states, with the canting angle $\phi_{{\rm SUB}n-n}$
chosen to minimize the energy $E_{{\rm SUB}n-n}(\phi)$, as shown in
Fig.~\ref{angleVSj2}. We also show separately the extrapolated
(SUB$\infty$) results obtained from Eq.~(\ref{Extrapo_E}) using the
data set $n=\{2,4,6\}$ shown. As is expected from our previous
discussion the energy curves themselves show very little evidence of
the phase transition at $\kappa = \kappa_{c_{1}}$, with the energy and
its first derivative seemingly continuous for both the spin-1 and
spin-$\frac{3}{2}$ cases.

\begin{figure*}[t]
\mbox{
   \subfloat[$s=1$\label{M_spin1}]{\scalebox{0.3}{\epsfig{file=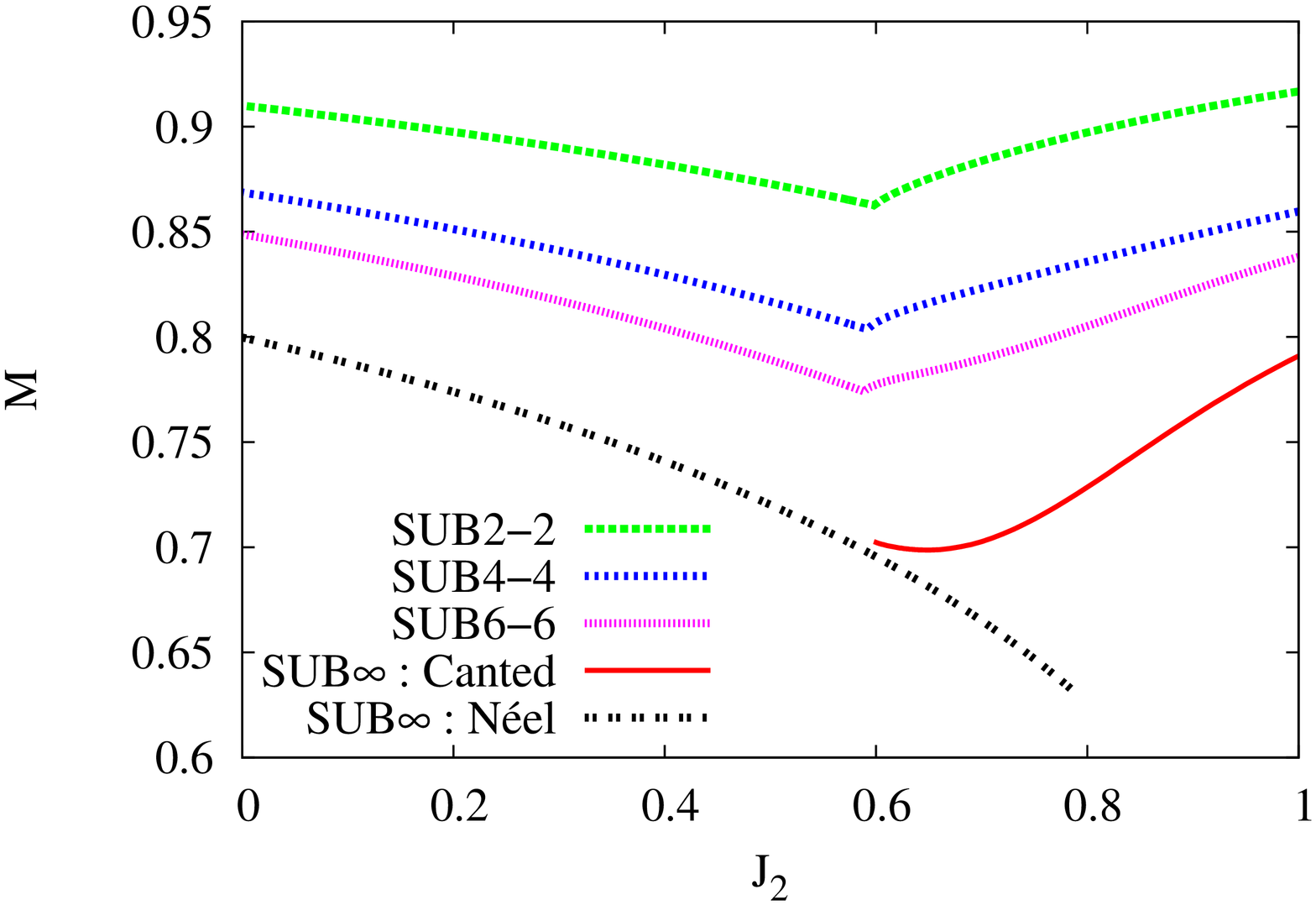,angle=270}}}
}
\mbox{
   \subfloat[$s=3/2$\label{M_spin_3Half}]{\scalebox{0.3}{\epsfig{file=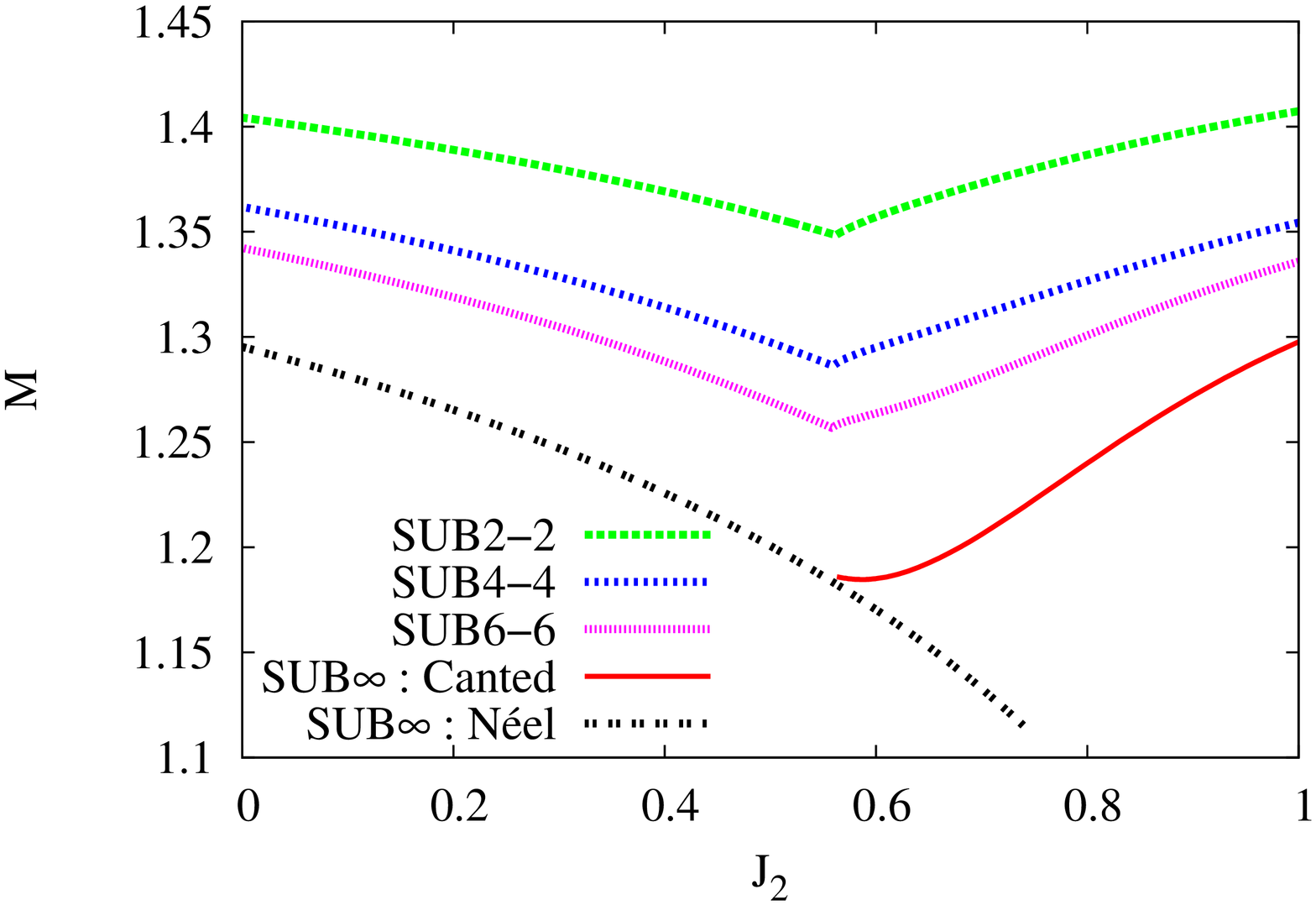,angle=270}}} 
}
\caption{Ground-state magnetic order parameter (i.e.,~the average
  on-site magnetization) versus $J_{2}$ for the N\'{e}el and canted
  phases of the (a) spin-1 and (b) spin-$\frac{3}{2}$ Union Jack
  Hamiltonian of Eq.\ (\ref{H}) with $J_{1} \equiv 1$. The CCM results
  using the canted model state are shown for various SUB$n$-$n$
  approximations with $n=\{2,4,6\}$, with the canting angle
  $\phi=\phi_{{\rm SUB}n-n}$ that minimizes $E_{{\rm
      SUB}n-n}(\phi)$. We also show the $n \rightarrow \infty$
  extrapolated result from using Eq.\ (\ref{Extrapo_M}).}
\label{M}
\end{figure*}

The transition between the N\'{e}el and canted phases is observed much
more clearly in our corresponding results for the gs magnetic order
parameter $M$ (the average on-site magnetization) shown in
Fig.~\ref{M}.
For the raw SUB$n$-$n$ data we display the results for the N\'{e}el
phase only for values of $\kappa < \kappa$$^{{\rm SUB}n-n}_{c_{1}}$
for clarity. However, the extrapolated (SUB$\infty$) results for the
N\'{e}el phase are shown for a wider range of values of $\kappa$,
using the extrapolation scheme of Eq.~(\ref{Extrapo_M}) and the
SUB$n$-$n$ results based on the N\'{e}el model state. For the canted
phase (for which $\phi_{{\rm SUB}n-n} \neq 0$) we can clearly only
show the extrapolated (SUB$\infty$) results using
Eq.~(\ref{Extrapo_M}), for regions of $\kappa$ for which we have data
for all of the set $n=\{2,4,6\}$. We see from Table~\ref{table_CritPt}
that we are limited (by the SUB2-2 results) to values $\kappa >
\kappa^{{\rm SUB}2-2}_{c_{1}} \approx 0.598$ for the $s=1$ case, and
$\kappa > \kappa^{{\rm SUB}2-2}_{c_{1}} \approx 0.563$ for the
$s=\frac{3}{2}$ case. From the data shown in Fig.~\ref{M}(a) for the
case of $s=1$, simple extrapolations of the SUB$\infty$ curve to lower
values of $\kappa < \kappa^{{\rm SUB}2-2}_{c_{1}} \approx 0.598$ using
simple spline fits in $\kappa$ give a corresponding estimate of
$\kappa_{c_{1}} \approx 0.580 \pm 0.015$ at which the N\'{e}el and
canted phases meet. The SUB$n$-$n$ extrapolations yield a nonzero
value for the average on-site magnetization of $M \approx 0.707 \pm
0.003$ at the phase transition point $\kappa_{c_{1}}$. Similarly for
the case of $s=\frac{3}{2}$ in Fig.~\ref{M}(b), simple extrapolations
of the SUB$\infty$ curve to lower values of $\kappa < \kappa^{{\rm
    SUB}2-2}_{c_{1}} \approx 0.563$ using simple spline fits in
$\kappa$ give a corresponding estimate of $\kappa_{c_{1}} \approx
0.535 \pm 0.005$ at which the N\'{e}el and canted phases meet. The
SUB$n$-$n$ extrapolations yield a nonzero value for the average
on-site magnetization of $M \approx 1.192 \pm 0.002$ at the phase
transition point $\kappa_{c_{1}}$. Thus the evidence from the
behaviour of the order parameter is that the transition at
$\kappa_{c_{1}}$ is a first-order one for both the cases $s=1$ and
$s=\frac{3}{2}$, in the sense that the order parameter does not go to
zero at $\kappa_{c_{1}}$, although it is certainly continuous at this
point. The extrapolated curves also provide every indication that the
derivative of the order parameter as a function of $\kappa$ is also
continuous (or very nearly so) at $\kappa = \kappa_{c_{1}}$ for both
the cases $s=1$ and $s=\frac{3}{2}$.

\begin{figure*}[t]
\mbox{
   \subfloat[$s=1$]{\scalebox{0.3}{\epsfig{file=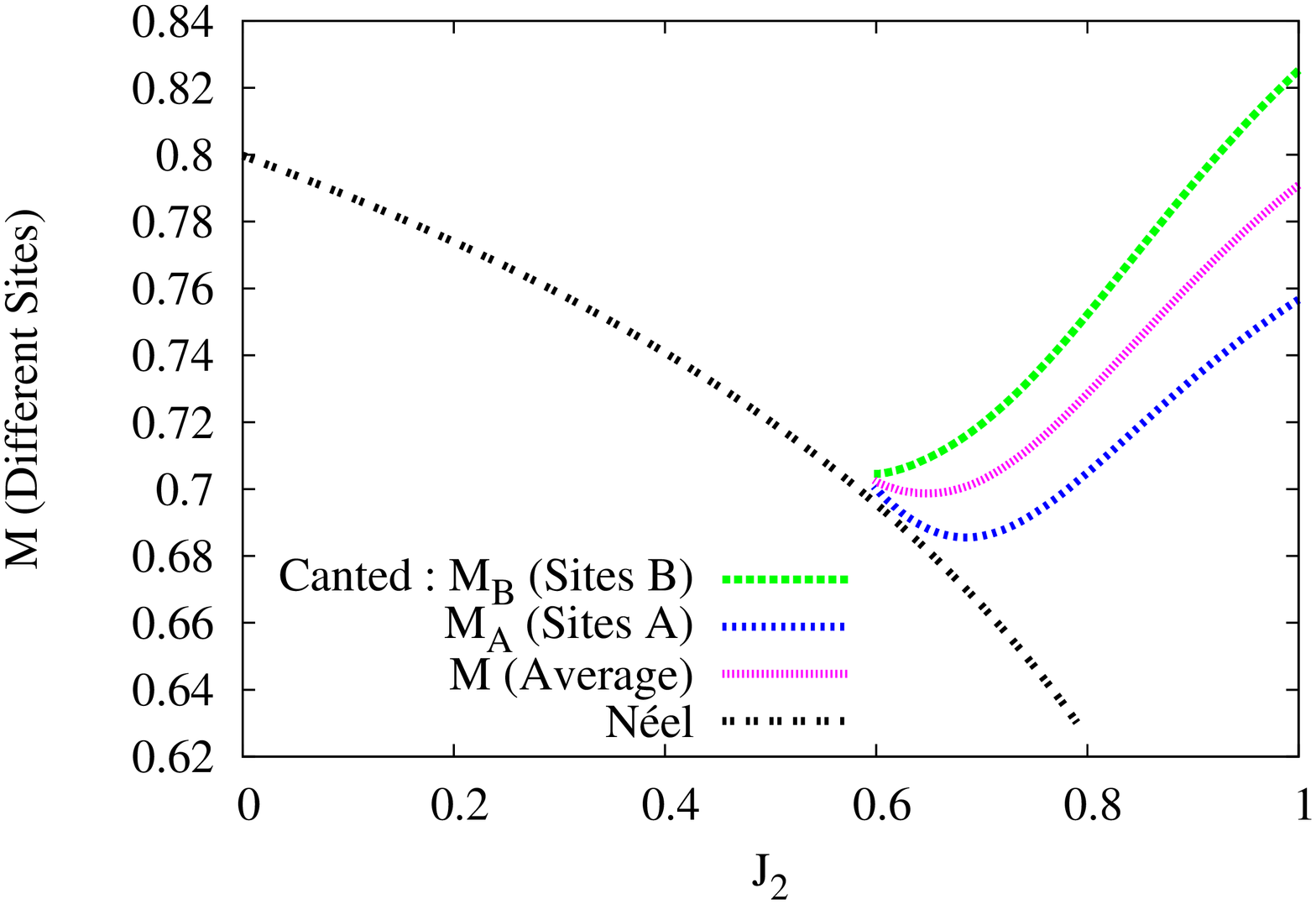,angle=270}}}
}
\mbox{
   \subfloat[$s=3/2$]{\scalebox{0.3}{\epsfig{file=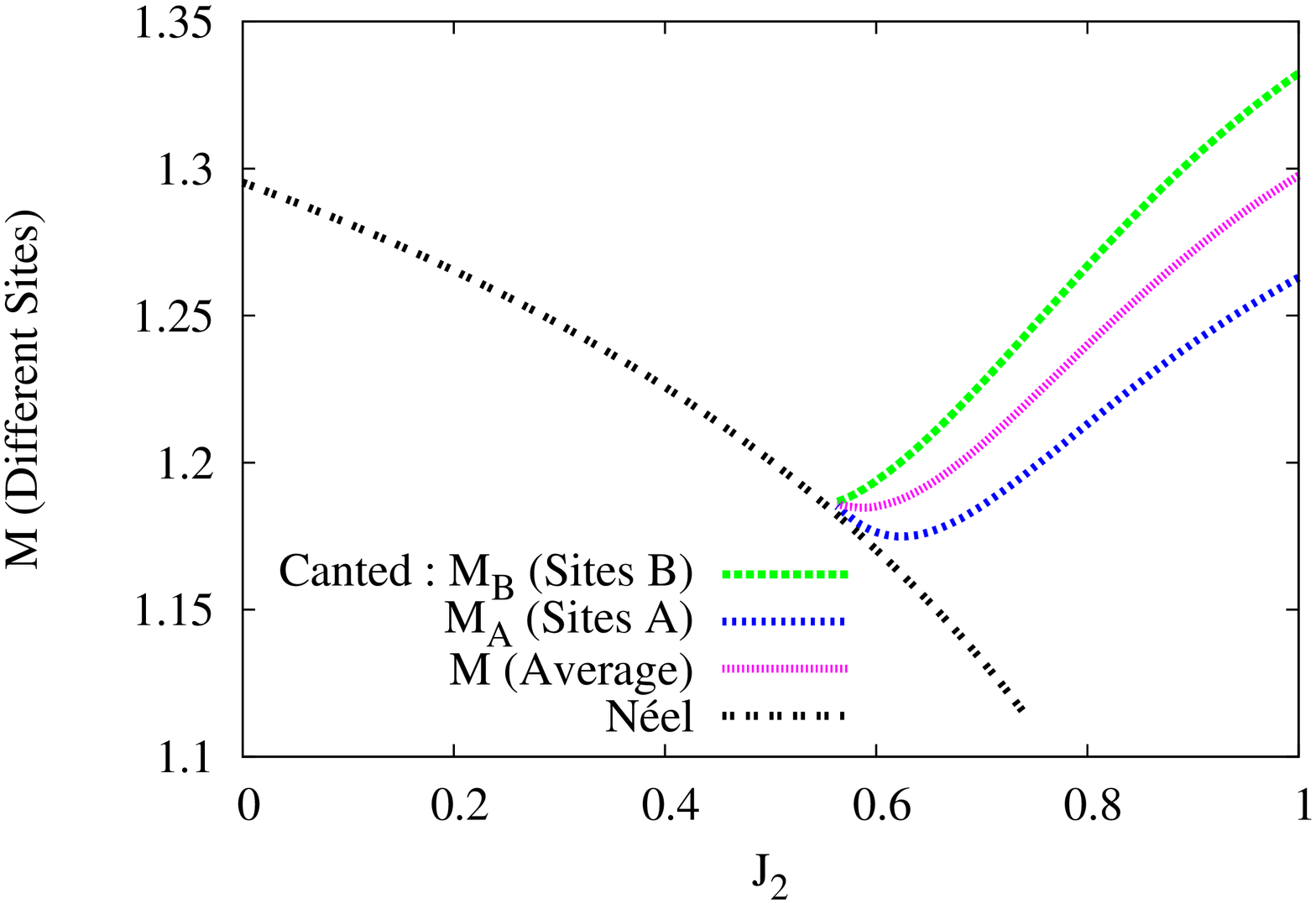,angle=270}}} 
}
\caption{Extrapolated curves (SUB$\infty$) for the ground-state
  magnetic order parameters (i.e., the on-site magnetizations)
  $M_{{\rm A}}$ at sites A (joined by eight bonds to other sites) and
  $M_{{\rm B}}$ at sites B (joined by four bonds to other sites) of
  the Union Jack lattice [and see Fig.~\ref{model}(a)] versus
  $J_{2}$ for the N\'{e}el and canted phases of the (a) spin-1 and (b)
  spin-$\frac{3}{2}$ Union Jack Hamiltonian of Eq.\ (\ref{H}) with
  $J_{1} \equiv 1$. The CCM results using the canted model state are
  shown for various SUB$n$-$n$ approximations ($n=\{2,4,6\}$) with the
  canting angle $\phi=\phi_{{\rm SUB}n-n}$ that minimizes $E_{{\rm
      SUB}n-n}(\phi)$.}
\label{M_DiffSites}
\end{figure*}

In Fig.~\ref{M_DiffSites}
we also show the corresponding extrapolated (SUB$\infty$) results for
the average on-site magnetization as a function of $J_{2}$ (with
$J_{1} \equiv 1$), or hence equivalently as a function of $\kappa$,
for both the A sites ($M_{{\rm A}}$) and the B sites ($M_{{\rm B}}$)
of the Union Jack lattice. We recall that, as shown in
Fig.~\ref{model}, each of the A and B sites is connected to four NN
sites on the square lattice by $J_{1}$ bonds, whereas each of the A
sites is additionally connected to four NNN sites on the square
lattice by $J_{2}$ bonds. The extrapolations are shown in exactly the
same regions, and for the same reasons, as those shown in
Fig.~\ref{M}.

We also comment briefly on the large-$J_{2}$ behaviour of our results
for the canted phase. (We note that for computational purposes it is
easier to re-scale the original Hamiltonian of Eq.~(\ref{H}) by
putting $J_{2} \equiv 1$ and considering small values of $J_{1}$.) For
the case of $s=\frac{1}{2}$ which we studied
previously~\cite{ccm_UJack_asUJ_2010}, the most interesting feature of
the CCM results using the canted state as model state is that in all
LSUB$n$ approximations with $n > 2$ a termination point $\kappa^{{\rm
    LSUB}n}_{t}$ is reached, beyond which no real solution can be
found. For example, we found~\cite{ccm_UJack_asUJ_2010} the values
$\kappa^{{\rm LSUB}4}_{t} \approx 80$ and $\kappa^{{\rm LSUB6}}_{t}
\approx 80$. This is a first indication that the canted state becomes
unstable at very large values of $\kappa$ against the formation of
another (as yet unknown) state, as we discuss further in
Sec.~\ref{Canted_vs_semiStripe} below. By contrast, in the present
cases of $s=1$ and $s=\frac{3}{2}$, the canted state shows no sign of
any instability for any SUB$n$-$n$ approximation with $n=\{2,4,6\}$ for
all values of $\kappa \leq 100$ studied.

Simple extrapolations to the $\kappa \rightarrow \infty$ limit of the gs
energy using the data at $\kappa \leq 100$ show that at large $J_{2}$
values we have $E/N \rightarrow -1.1652J_{2}$ for $s=1$ and $E/N
\rightarrow -2.4943J_{2}$ for $s=\frac{3}{2}$. These numerical
coefficients are almost exactly half of the values quoted in
Table~\ref{table_EandM_results} for the case $J_{2}=0$. This is as
expected since both the $\kappa \rightarrow 0$ and the $\kappa
\rightarrow \infty$ limits of the Union Jack model are the
square-lattice HAF, where in the latter case the square lattice
contains only half the original sites, namely the A sites. Similarly,
the extrapolated SUB$\infty$ values at larger values of $\kappa=100$
for the on-site magnetization on the A sites are $M_{{\rm A}}
\rightarrow 0.807$ for $s=1$ and $M_{{\rm A}} \rightarrow 1.303$ for
$s=\frac{3}{2}$. These values are again in very good agreement with
those shown in Table~\ref{table_EandM_results} for the $J_{2} = 0$
limit. The corresponding asymptotic values for the B-site
magnetization are consistent with $M_{{\rm B}} \rightarrow 1.0$ and
$M_{{\rm B}} \rightarrow 1.5$, as expected for large values of
$J_{2}$.

\subsection{Canted state versus the semi-striped state}
\label{Canted_vs_semiStripe}
For the Union Jack model considered here, but for the case
$s=\frac{1}{2}$ studied previously, we
predicted~\cite{ccm_UJack_asUJ_2010} a second phase transition at
$\kappa=\kappa_{c_{2}} \approx 125 \pm 5$ using the same CCM technique
as used here. At this upper transition the ferrimagnetic quantum
canted phase shown in Fig.~\ref{model}(a) gives way to the quantum
ferrimagnetic semi-striped phase shown in Fig.~\ref{model}(b), such
that for value $\kappa > \kappa_{c_{2}}$ the semi-striped phase
becomes lower in energy. In order to investigate the possible similar
stabilization of the quantum semi-striped state at high values of the
frustration parameter $\kappa$, we have performed similar CCM
calculations here, for the $s=1$ and $s=\frac{3}{2}$ cases, based on
the semi-striped state as model state.

As already noted above, we find no sign of instability in the canted
state, for either of the cases $s=1$ or $s=\frac{3}{2}$, for any level
of SUB$n$-$n$ approximation with $n \leq 6$ at all values $\kappa \leq
100$ investigated, unlike for the corresponding LSUB$n$ approximations
in the $s=\frac{1}{2}$ case that terminated in this range for all
values $2 < n \leq 6$ investigated. Furthermore, we find no evidence
at all that the semi-striped phase has lower energy than the canted
phase for any value of $\kappa$ for either of the case $s=1$ or
$s=\frac{3}{2}$. Nonetheless, simple extrapolations to the $\kappa
\rightarrow \infty$ limit of the gs energy of the semi-striped phase,
using the SUB$n$-$n$ data with $n=\{2,4,6\}$ at $\kappa \leq 1000$,
show that at large $J_{2}$ values we have $E/N \rightarrow -1.1646
J_{2}$ for $s=1$ and $E/N \rightarrow -2.4939J_{2}$ for
$s=\frac{3}{2}$ (with $J_{1} \equiv 1$). The corresponding asymptotic
($\kappa \rightarrow \infty$) values for the average gs on-site
magnetization of the semi-striped state are $M_{\rm A} \rightarrow
0.807$ on the A sites and $M_{\rm B} \rightarrow 1.303$ on the A sites
and $M_{\rm B} \rightarrow 1.5$ on the B sites for the
$s=\frac{3}{2}$ case, just as for the corresponding large-$\kappa$
limits of the canted state, and as corresponds to the pure
square-lattice HAF as already noted above.

\section{Discussion and conclusions}
\label{discussion}
In an earlier paper~\cite{ccm_UJack_asUJ_2010} we used the CCM to
study the effect of quantum fluctuations on the zero-temperature gs
phase diagram of a frustrated spin-$\frac{1}{2}$ Heisenberg
antiferromagnet (HAF) defined on the 2D Union Jack lattice of
Fig.~\ref{model}. In the present paper we have extended the analysis
to the two computationally more challenging cases where all the
lattice spins have spin quantum number either $s=1$ or
$s=\frac{3}{2}$. We have once again concentrated on the case where the
NN $J_{1}$ bonds are antiferromagnetic ($J_{1} > 0$) and the competing
NNN $J_{2} \equiv \kappa J_{1}$ bonds in the Union Jack array have a
strength in the range $0 \leq \kappa < \infty$. On the underlying
bipartite square lattice there are thus two types of sites, viz., the
A sites that are connected to the four NN sites on the B sublattice
with $J_{1}$ bonds and to the four NNN sites on the A sublattice with
$J_{2}$ bonds, and the B sites that are connected only to the four NN
sites on the A sublattice with $J_{1}$ bonds. The $\kappa = 0$ limit
of the model thus corresponds to the HAF on the original square
lattice (of both A and B sites), while the $\kappa \rightarrow \infty$
limit corresponds to the HAF on the square lattice comprised of only A
sites. We have seen that at the classical level (corresponding to the
case where the spin quantum number $s \rightarrow \infty$) this Union
Jack model has only two stable gs phases, one with N\'{e}el order for
$\kappa < \kappa$$^{{\rm cl}}_{c} = 0.5$ and another with canted
ferrimagnetic order for $\kappa > \kappa$$^{{\rm cl}}_{c}$. We have
therefore first used these two classical states as CCM model states to
investigate the effects of quantum fluctuations on them.

For the spin-1 model we find that the phase transition between the
N\'{e}el antiferromagnetic phase and the canted ferrimagnetic phase
occurs at the value $\kappa_{c_{1}} = 0.580 \pm 0.015$, whereas for
the spin-$\frac{3}{2}$ model we find that the phase transition occurs
at $\kappa_{c_{1}} = 0.545 \pm 0.015$. The evidence from our
calculations is that the transition at $\kappa_{c_{1}}$ is a subtle
one. From the energies of the two phases it appears that the
transition is second-order, as in the classical case. However, on
neither side of the transition at $\kappa_{c_{1}}$ does the order
parameter $M$ (i.e., the average on-site magnetization) go to
zero. Instead, as $\kappa \rightarrow \kappa_{c_{1}}$ from either
side, $M \rightarrow 0.707 \pm 0.003$ for the $s=1$ case and $M
\rightarrow 1.192 \pm 0.002$ for the $s=\frac{3}{2}$ case, which are
more indicative of a first-order transition. Furthermore, the slope
$dM/d\kappa$ of the average on-site magnetization as a function of
$\kappa$ also seems to be either continuous or to have only a very
weak discontinuity at $\kappa = \kappa_{c_{1}}$ in both of the cases
$s=1$ and $s=\frac{3}{2}$.

In the case of the previously studied spin-$\frac{1}{2}$
model~\cite{ccm_UJack_asUJ_2010} we found evidence for a second
quantum phase transition at a value $\kappa = \kappa_{c_{2}} \approx
125 \pm 5$ at which the quantum ferrimagnetic canted phase for $\kappa
< \kappa_{c_{2}}$ yields to a lower-energy quantum ferrimagnetic
semi-striped phase for $\kappa > \kappa_{c_{2}}$. Our LSUB$n$ results
with $n > 2$ for the spin-$\frac{1}{2}$ model gave clear evidence that
the canted phase terminated at some value $\kappa^{{\rm LSUB}n}_{t}$
such that for $\kappa > \kappa^{{\rm LSUB}n}_{t}$ no real solution
based on the canted state as model state existed. These termination
points were a preliminary signal of the actual phase transition at
$\kappa_{c_{2}}$. In neither the $s=1$ nor $s=\frac{3}{2}$ cases
considered here do we find any corresponding upper termination points
for any SUB$n$-$n$ approximation based on the canted state as model
state, for any value of $n \leq 6$ and for any value of $\kappa$ up to the
highest value $\kappa = 100$ investigated. Furthermore, for all
approximation and for all values of $\kappa$ investigated the
semi-striped phase always lies higher in energy than the quantum
canted phase for both the $s=1$ and $s=\frac{3}{2}$ cases, unlike what
was found~\cite{ccm_UJack_asUJ_2010} for the $s=\frac{1}{2}$
case. Thus, our results clearly show that only the $s=\frac{1}{2}$
model has a second quantum phase transition at $\kappa=\kappa_{c_{2}}$
between the two ferrimagnetic (canted and semi-striped) phases. No
such transition is predicted to occur at a finite value of $\kappa$
for either of the $s=1$ or $s=\frac{3}{2}$ cases.

To the best of knowledge no other studies of the Union Jack model for
lattice spins with spin quantum number $s>\frac{1}{2}$ have been
performed, and hence we have no other results with which to
compare. Nevertheless, the consistency of the present results and
those of our previously studied spin-$\frac{1}{2}$
case~\cite{ccm_UJack_asUJ_2010} give credence to our
results. Furthermore, as has been noted elsewhere \cite{Bi:2008_JPCM},
high-order CCM results of the sort presented here have been seen to
provide very accurate and reliable results for a wide range of such
highly frustrated spin-lattice models. Many previous applications of
the CCM to unfrustrated spin models have given excellent quantitative
agreement with other numerical methods (including exact
diagonalization (ED) of small lattices, quantum Monte Carlo (QMC), and
series expansion (SE) techniques). A typical example is the HAF on the
square lattice, which is the $\kappa=0$ limit of the present
model~\cite{ccm_UJack_asUJ_2010}. We have compared our own results for
this $\kappa=0$ case with both SWT and SE results, both for the
spin-$\frac{1}{2}$ case in our earlier work~\cite{ccm_UJack_asUJ_2010}
and for the spin-1 case here, and shown that there is excellent
agreement in both cases. This adds further credence to the validity of our
results at nonzero values of $\kappa$.

\section*{Acknowledgment}
We thank D.J.J.~Farnell for fruitful discussions.

\end{document}